\RequirePackage{fix-cm}
\documentclass[smallextended]{svjour3}       
\smartqed  
\usepackage{graphicx}
\usepackage{hyperref}
\usepackage[colorinlistoftodos]{todonotes} 
\usepackage{xcolor} 
\usepackage{algorithm}
\usepackage[noend]{algpseudocode}
\usepackage{amsmath}

\begin{document}

\title{Measuring the \emph{Salad Bowl}: Superdiversity on Twitter \thanks{This work has been funded by the Horizon2020 European projects ``SoBigData Research Infrastructure - Big Data and Social Mining Ecosystem" (grant agreement 654024), and ``HumMingBird  -- Enhanced migration measures from a multidimensional perspective" (grant agreement no 871042).}
}


\author{Laura Pollacci         \and
        Alina S\^irbu  \and
        Fosca Giannotti \and
        Dino Pedreschi
}


\institute{L. Pollacci, A. S\^irbu, D. Pedreschi\at
              Department of Computer Science, University of Pisa \\
              Largo B. Pontecorvo 3, 56127 Pisa, Italy\\
              \email{laura.pollacci@di.unipi.it, (name.surname)@unipi.it}           
           \and
          F. Giannotti \at
Scuola Normale Superiore, P.za dei Cavalieri, 7, 56126 Pisa PI\\ \email{fosca.giannotti@sns.it}}


\maketitle

\begin{abstract}
Superdiversity refers to large cultural diversity in a population due to immigration. In this paper we introduce a \emph{superdiversity index} based on the changes in the emotional content of words used by a multi-cultural community, compared to the standard language. To compute our index we use Twitter data and we develop an algorithm to extend a dictionary for lexicon based sentiment analysis. We validate our index by comparing it with official immigration statistics available from the European Commission's Joint Research Center, through the D4I data challenge. We show that, in general, our measure correlates with immigration rates, at various geographical resolutions. Our method produces very good results across languages, being tested here both on English and Italian tweets. We argue that our index has predictive power in regions where exact data on immigration is not available, paving the way for a nowcasting model of immigration rates.

\end{abstract}

\section{Introduction}


Superdiversity is a relatively new term that refers to a major cultural diversity in the population due to recent migration phenomena~\cite{vertovec2007super}. Superdiversity is difficult to measure in a population. In terms of migration, one can measure diversity based on the languages spoken in a country~\cite{moise2016tracking}, immigrant percentage, and other criteria, however to our knowledge there is no established superdiversity index. At the same time, there is a very close interplay between diversity, migrant integration, migrant attachment to the home country, making the measurement of diversity and superdiversity far from straightforward.  

In this work we propose a novel measure of superdiversity that we call the Superdiversity Index ($SI$). This is based on the emotional content of words in a community. Persons with different cultural backgrounds will necessarily associate different emotional valences to the same word. Therefore, a multi-cultural community will display a use of the local language that is different in its emotional content compared to a standard expected use. We believe that a more diverse community has a larger distance between standard and actual emotional valences. 

Our $SI$ is built on Twitter data and lexicon based sentiment analysis. Specifically, we introduce an algorithm able to calculate emotional valences for words used on Twitter by various communities, that we apply to the local language. We then compare the calculated valences with emotional valences from a standard tagged lexicon. The distance between the two gives a measure of diversity. We compute our $SI$ at different geographical resolutions, for the United Kingdom (UK) and Italy, and we show that $SI$ values correlate well with foreign immigration rates. Furthermore, our $SI$ outperforms by far other possible measures of (super)diversity extracted from the same Twitter data, such as the use of multiple languages or lexical richness. 

An important characteristic of our $SI$ is the high correlation with immigration rates. This prompts us to expect that it can become an important feature in a nowcasting model of migration stocks. Migration flows and stocks are typically measured at national level by official statistics offices, for example through regular population censuses. The same applies for other effects such as social and economic integration. These data are critical for development of policies to optimise the beneficial effects of migration under all criteria. However, due to their nature, they can become outdated (censuses are organised every ten years), or information can be inconsistent when moving from one national statistics office to another. Hence, lately, alternative data including social media data are starting to be proposed to measure migration effects in various settings~\cite{zagheni2012you,sirbu2019,kim2020digital}. Our work is a first step towards nowcasting immigration from Twitter data.

The rest of this paper is organised as follows.  Section~\ref{sec:introSI} introduces $SI$, a novel superdiversity index, including a high-level description of the algorithm used to estimate emotional valences of words used by a community. Section~\ref{sec:lexical_res} presents the lexical resources employed to estimate emotional valences of words. A detailed description of the algorithm used to estimate emotional valences, the Emotional Spreading Algorithm (EmoSA), is provided in Section~\ref{sec:emosa}. We describe the datasets used for the analysis and the preprocessing steps in  Section \ref{subsec:untagged_dataset} and \ref{subsec:tagged_dataset}, while in Section \ref{sec:word_eval} we evaluate the emotional valences obtained with EmoSA. The detailed description of the procedure to compute  the $SI$, and the values obtained are shown in Section~\ref{sec:SI_eval}, and compared  with immigration rates. Section~\ref{sec:discussion} discusses the possible application of this work for nowcasting immigration rates, while Section~\ref{sec:conclusions} concludes the paper, after presenting the related work in Section~\ref{sec:related}.

\section{Introducing the Superdiversity Index}
\label{sec:introSI}

We start from the hypothesis that different cultures associate different emotional valences to the same word. So, a culturally diverse community will show a use of the language that is different from a standard expected use. Thus, we defined our Superdiversity Index ($SI$) as the \emph{distance} between the \emph{community-dependent} and the \emph{standard} emotional valences for a set of words.

\begin{figure}
    \centering
    \includegraphics[width=.75\textwidth]{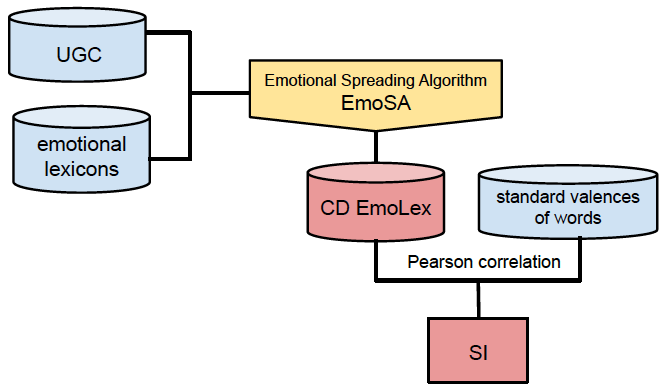}
    \caption{Overall process to compute SI.}
    \label{fig:siProcess}
\end{figure}

The overall process to calculate $SI$ for a certain community is summarised in Figure~\ref{fig:siProcess}.
The first step is to compute the community-dependent emotional lexicon (CD EmoLex), i.e. a list of words labelled with an emotional polarity value determined by the community. This is obtained by applying the Emotional Spreading Algorithm (EmoSA) (see Section \ref{sec:emosa} and~\cite{pollacci2017sentiment} for details). The algorithm takes as input a user-generated content (UCG) dataset, in our case geolocalised tweets, and a collection of existing labelled emotional lexicons. The CD EmoLex returned contains emotional values that refer to the specific emotional usage of words by the community of users posting on Twitter from the same geographical area. 

The emotional valences in the CD EmoLex are compared to the standard emotional valences from an established lexicon. The reliability of the standard emotional lexicon is crucial since we consider it as the \emph{ground truth} of the emotional usage of words. For this reason, we employ ANEW~\cite{bradley1999affective}, a well-known emotional lexicon developed and distributed by NIMH Center for Emotion and Attention (CSEA) at the University of Florida (see Section~\ref{sec:lexical_res}). We compute the Pearson correlation ($r$) between the standard and the CD EmoLex as a measure of similarity. We repeat the whole procedure several times, and then we calculate the average of the Pearson correlations ($\bar{r}$). Finally, we compute the $SI$ as 

\begin{equation}
\label{eq:si}
SI=\frac{1-\bar{r}}{2}
\end{equation}
 An $SI$ value of $0$ corresponds to no diversity, i.e. emotional content identical between the CD EmoLex and the standard lexicon. A value of $0.5$ corresponds to no correlation between the two lexicons, i.e. the emotional content of words used by the community on Twitter is not related to the standard emotional content. A value of $1$, which is very unlikely, would correspond to the use of terms with the opposite emotional content compared to standard.

 The proposed $SI$ inevitably includes a component related to the performance of the EmoSA algorithm. As explained in Section \ref{sec:emosa}, EmoSA provides an estimate based on a subset of tweets coming from a subset of the total community to be analysed, and is based on several assumptions. Thus, this estimate unavoidably includes some error. In fact, the correlation $r$, and in consequence the value of $SI$, depends both on the different use of the language and on the error of the algorithm. Even so, we believe that the $SI$ we propose can be efficient in quantifying diversity. This because the error component should be stable when changing the cultural mix of the community analysed, or at most it can increase as multi-culturality increases, so it does not affect negatively the relation between the $SI$ and diversity, but may even enhance it.

\section{Lexical Resources and Data}
\label{sec:lexical_res}

To apply EmoSA (see section \ref{sec:emosa}) and calculate $ SI $, we take advantage of several resources, which include, lists of lemmas tagged with sentiment values, Twitter data and immigration data. 

\subsection{Lexical Resources}
The reliability of the tagged lexicon is crucial for our algorithm. Among the several prominent polarity lexicons attested in the literature, such as Sentiment Orientation CALculator (SO-CAL) \cite{taboada2011lexicon}, Sentiment Treebank \cite{socher2013recursive},  MPQA (Multi-Perspective Question Answering) Subjectivity Lexicon (MPQA SL) \cite{riloff2003learning,wilson2005recognizing}, General Inquirer \cite{stone1966general}, SentiWordNet \cite{esuli2007sentiwordnet}, AFINN \cite{nielsen2011new}, and Opinion Lexicon \cite{hu2004mining,liu2005opinion} we chose ANEW \cite{bradley1999affective}. 
The \emph{Affective Norms for English Words} (ANEW) lexicon provides a set of normative emotional ratings for 1,034 words in terms of \emph{pleasure}, \emph{arousal}, and \emph{dominance}\footnote{Pleasure represents positive versus negative emotions. Arousal finds the two extremes of the scale of values in ``calm" and ``exciting". The dominance dimension determines if the subject feels in control of the situation or not. Each dimension ranges in $[0,10]$.}.

ANEW, as well as AFINN, MPQA SL, and SO-CAL, and unlike, e.g., SentiWordnet, is a manually tagged lexicon. This means that the polarity annotations tend to be more accurate and reliable. 
For our algorithm, we used the \emph{pleasure} dimension as a sentiment valence, as evaluated by both male and female subjects.
Furthermore, the words in ANEW are tagged following a continuous scale of values that allows us to capture different nuances between the polarity values. Lexicons such as MPQA SL, Sentiment Treebank, and  General Inquirer includes not-continuos polarity scales. In the MPQA SL lemmas are tagged as positive and negative with a further annotation for intensity  (strong,  weak); Sentiment Treebank proposes a five-class classification (\emph{very positive}, \emph{positive}, \emph{neutral}, \emph{negative}, and \emph{very negative}); General Inquirer and Opinion Lexicon classify lemmas as positive or negative. Moreover, unlike the manually tagged lexicon SO-CAL in which sentiment label is an integer in $[-5, +5]$, ANEW also includes objective words, and thus neutral, words. Finally, as for instance SO-CAL and SentiWordNet, ANEW assigns PoS tags to lemmas, unlike Opinion Lexicon and AFINN.

\emph{SentiWordNet} is a publicly available lexical resource. For each lemma in the lexicon there are three numerical scores (\emph{Obj(s)}, \emph{Pos(s)} and \emph{Neg(s)}) describing how objective, positive, and negative the lemma is.
In order to compute a unique polarity score for each lemma, we adopt the difference between the positive and negative score as an overall sentiment valence, properly scaled to the interval [0, 10] \cite{guerini2013sentiment}. Since polarities' distribution is strongly heterogeneous we balanced the obtained lexicon by choosing the number (\emph{n}) of items in the least represented class (negative) and selecting from other classes the \emph{n} most strongly polarized lemmas.

The \emph{Full List of Bad Words Banned by Google} of the ``What Do You Love'' (WDYL) Google project is a list of 550 swear and curse terms\footnote{The service is now inactive. The list can be downloaded from Free Web Header: \href{https://www.freewebheaders.com/full-list-of-bad-words-banned-by-google/}{Full List of Bad Words Banned by Google}}. We manually enriched each word with a 0.0 valence score, hence these terms are considered strongly negative.

Due to the multi-language nature of our experiment, we translate each of the three English lexicons above into the Italian language. To obtain the most accurate translation we combine and cross-check two different API services: Googletrans\footnote{\url{https://pypi.org/project/googletrans/}} and Goslate\footnote{\url{https://pypi.org/project/goslate/}}. In addition, we also impose some constraints, such a threshold on the translation confidence score. Finally, in order to be more confident in the translation, we select only lemmas attested in the PAISA'\cite{lyding2014paisa} corpus, which contains about two hundred and fifty million tokens labelled with their frequency, retrieved from about three hundred and eighty thousand Italian web texts.

\subsection{Untagged Twitter Dataset}
\label{subsec:untagged_dataset}
The untagged Twitter dataset is a subset of the one used in \cite{coletto2017perception} and it is composed of just under 73,175,500 geolocalised tweets gathered for 3 months, from the 1st August to the 31st October of 2015. From these, we have selected only the tweets originating in the UK or Italy. 
They have been preprocessed to obtain cleaned tweets only coming from the United Kingdom and Italy. The first problem faced is origin and language selection, which was accomplished by using the metadata of the tweets themselves. Specifically, for the language, we used the value provided by the Twitter API, while for the geolocation we used the geographical coordinates. Thus we selected only the tweets that contained the geographical coordinates, and did not include those containing only the \emph{place}, since this is not reliable information on the source of a tweet. To obtain texts processable in an effective way by automatic methods, we employ both rule-based and general-purpose NLP pipeline. Then, all tweets are lemmatised and parts of speech are tagged using the POS tagger TreeTagger \cite{schmid1994probabilistic,schmid1995improvements}. Finally, to reduce the noise we selected only nouns, adjectives, and verbs. This selection allowed us to obtain only significant words from the sentiment and meaning point of view. At the end of the preprocessing phase, we obtained two cleaned and standardised subsets: one composed of all the English tweets posted from the United Kingdom, and one composed of all the Italian tweets posted from Italy.

A second preprocessing stage had as objective the assignment of tweets from each dataset to a city in the UK or Italy. This was required to match locations in D4I (see~\ref{sec:d4i}) to our results. Each tweet has associated a location, including the city or town it came from. Thus, for each tweet, we check if its origin is attested in D4I. If the city is found, we assign to the tweet the related NUTS\footnote{The Classification of Territorial Units for Statistics (NUTS), in French ``\emph{Nomenclature des unit\'es territoriales statistiques}'', is a standard geocode referring the subdivisions of countries for statistical purposes. The \href{http://ec.europa.eu/eurostat/web/nuts/history}{standard} is developed and regulated by the European Union. } code. Otherwise, we perform a dedicated rule-based pipeline. This step is often required since in D4I locations are at the city level, while several tweet origins are at the district or town level. 
First, the MediaWiki API\footnote{MediaWiki API main page: \href{https://www.mediawiki.org/wiki/API:Main_page}{MediaWiki}} is exploited by using the tweet's location as a key-word. When the API allows us to extract the city referred to the tweet's origin from the MediaWiki page's info-box, we assign it to the tweet. If not, the location value is used as a Google Search API\footnote{Google Custom Search API page: \href{https://developers.google.com/custom-search/?csw=1}{Google Custom Search}} parameter in order to extract URLs of the first five pages. From these, URLs pointing to Wikipedia are selected and, as before, are used to extract the city referred to by the location from the info-box. If the combination of the already mentioned APIs does not allow us to retrieve a city attested in the D4I dataset, we exploit the Google Maps API first and the Geopy\footnote{Geopy site: \href{https://pypi.org/project/geopy/}{Geopy}} python library then. These two are used after the first searching phase due to their rate and call limits, even though they are generally more accurate. All the retrieved cities are finally matched with those in the D4I dataset and each tweet is labeled with NUTS codes at three levels (\emph{NUTS1}, \emph{NUTS2}, and \emph{NUTS3}). Following this procedure, we are able to associate a NUTS code with over 94\% of the UK tweets and to over 97.5\% of the Italian tweets. 

It is, however, worth pointing out that in both Twitter datasets there are several tweets geo-labeled with the region/county or country name. Also, in particular in the Italian dataset, many tweets are labeled with multi-language city names translations, such as ``N\'{a}poles'', ``Trentino-Alto Adigio'', ``Venecia'', ``S\~{a}o Gimian'' and ``S\~{a}o Remo'', ``Anc\^{o}ne'', ``Naturns'' and ``Florencia''. Tweets belonging the first case are discarded due to the geographical multilevel nature of our analysis. In fact, country or region/county tags do not allow us to reach a fine-grained geographical level. Instead, tweets belonging to the second case are not ruled out a priori but they are rarely assigned a NUTS code by our pipeline.

The size and number of cities covered by the two datasets  after preprocessing are showed in Table \ref{tab:data}. 

\begin{table}
\centering
	\begin{tabular}{|l|c|c|}
    	\hline
        Dataset & \# tweets & \# matched cities\\
        \hline
        UK & 2,088,346 & 9,603  \\ 
        Italy & 274,885 & 6,050 \\
        \hline
	\end{tabular}
    \caption{Dataset details for the UK and Italy. }
\label{tab:data}
\end{table}

\subsection{Tagged Twitter Dataset}
\label{subsec:tagged_dataset}
This dataset is composed of 3,718 publicly available tweets and their sentiment classifications, retrieved from three different sources:

\emph{The Semeval 2013 Message Polarity Classification competition (task B)}\footnote{The Semeval 2013 Message Polarity Classification competition (task B), https://www.cs.york.ac.uk/semeval-2013/}. The original dataset consisted of  a 12-20K messages corpus on a range of topics, classified into positive, neutral and negative classes. We retrieved 2,752 such tweets, that were still available on the Twitter platform. These were passed through the language detection algorithm provided by the Python package Langdetect, to ensure they were written in English, leaving us with 2,547 tweets in the final dataset (428 negative, 1,347 neutral and 970 positive).

\emph{The Semeval 2014 Message Polarity Classification competition (task B)}\footnote{The Semeval 2014 Message Polarity Classification competition (task B), http://alt.qcri.org/semeval2014/}. Similar to Semeval 2013, this corpus consisted originally in 10000 tweets, out of which we downloaded 687 English tweets (142 negative, 319 neutral and 226 positive).

\emph{Earth Hour 2015 corpus}\footnote{The Earth Hour 2015 corpus: ~https://gate.ac.uk/projects/decarbonet/datasets.html}.
This dataset contains 600 tweets annotated with Sentiment information (Positive, Negative, Neutral) where each annotation is triply-annotated through a crowdsourcing campaign. 
Out of these, 370 tweets were still available for download through the Twitter platform (30 negative, 185 neutral and 80 positive).

We observe that a small proportion of tweets is tagged with a negative valence, with neutral tweets being most prevalent, in all three datasets. This can be seen as a characteristic of Twitter messages in general. This dataset will be used in the following to evaluate the performance of our community-dependent emotional lexicon, compared to an already established lexicon, in classifying tweet sentiment.
The tweets are normalised and lemmatised similar to the Untagged Twitter Dataset.

\subsection{D4I Dataset}
\label{sec:d4i}
The D4I dataset~\cite{d4i} contains the concentration of migrants in all cities of eight EU countries: Spain, Germany, Italy, France, Netherlands, Portugal, United Kingdom, and Ireland. Migrant counts are based on the 2011 EU population census.  Migrants are counted based on three different levels of aggregation: by country, continent and EU versus non-EU. 
For our analysis, we focused only on data for the UK and Italy and summing EU and non-EU immigrant counts to obtain total immigration levels, since diversity is caused by all migrant types.

\section{Computing emotional valences from Twitter with EmoSA}
\label{sec:emosa}

Lexicon-based sentiment analysis methods use a lexicon of words tagged with an emotional valence to assign sentiment to text, by aggregating the valences of single words. These lexicons can be tagged manually,  as in the case of ANEW (Section~\ref{sec:lexical_res}), or can be generated automatically. While manual annotations can be more accurate, lexicons can be too reduced to allow for accurate sentiment analysis, especially for small texts such as tweets. Here we introduce EmoSA, a method to extend the lexicon automatically starting from an existing seed one, using sentiment spreading on a co-occurrence network of terms. The value of our method is two-fold. First, it creates a much larger tagged lexicon (namely Community-Dependent Emotional Lexicon (CD EmoLex)) that can be used for lexicon-based sentiment analysis. Second, and most important for the definition of our $SI$, the emotional valences attached to words depend on the way words are used by the community, thus the resulting emotional valences can be an indication of cultural diversity. 

For a global vision, EmoSA is based on a starting text corpus, in our case an untagged Twitter dataset (UTD, see Section \ref{subsec:untagged_dataset}), and on the standard emotional lexicon ANEW (Section~\ref{sec:lexical_res}). The algorithm consists of two stages (see Algorithm~\ref{alg:emosa}). First we build a co-occurrence network of terms in the tweet dataset. Subsequently, we initiate the epidemic spreading of emotional valences starting from seed values. In the following we provide details for each phase.

To perform the EmoSA, we split the ANEW lexicon into two (50\% each) equal training ($TRAIN$) and test ($TEST$) datasets.
The terms in the training dataset are used for seeding EmoSA, while the test dataset is used to validate the algorithm and to compute the $SI$. 

\begin{algorithm}[t]
\caption{EmoSA.}\label{alg:emosa}
    \begin{algorithmic}[1]
        \Procedure{compute\_valences}{$TRAIN,TEST,UTD$}
        \State $V, N \gets build\_network(UTD)$ \Comment {build word network}
        \State $V^* \gets sentiment\_spreading(V,TRAIN,N) $
        \State $V^* \gets V^* \cap TEST$ \Comment{Select valences of words for the TEST dataset}
        \State \bf{return} $V^*$
        \EndProcedure
    \end{algorithmic}
\end{algorithm}

\subsection{Building the network}
Starting from an Untagged Twitter dataset (Section \ref{subsec:untagged_dataset}), we build a network of lemmas where each lemma corresponds to one node. Two nodes are connected by an edge if there is at least one tweet where both lemmas appear.

Hence, the network is an unweighted co-occurrence graph based on the target tweets to be classified. We use a large amount of tweets to build this network, thus we expect that lemmas with positive valence will be mostly connected to other positive lemmas, while those with negative valences will be connected among themselves. We consider only tweets not containing a negation (``don't", ``not", etc.), since with negations it is difficult to understand which lemmas from the negated tweet can be considered connected in the network, and which not.

\subsection{Sentiment Spreading}
Once the network of lemmas is obtained, we start to add valences to each node in the network. We start from a \emph{seed lexicon}, which is typically reduced in size. In the next section we will show results obtained when the seed ($TRAIN$) is 50\% of the ANEW lexicon (the other half is used to validate the results, i.e. $TEST$ data), together with all lemmas in the SentiWordNet and Bad words lexicon. This seed allows us to assign valences to a reduced number of nodes in the network. This is the initial state of our epidemic process.

Starting from the initial state, we follow a discrete time process where at each step sentiment valences spread through the network. At time $t$, for all nodes $i$ which do not have any valence, the set of neighbouring nodes $N(i)$ is analysed, and $i$ takes a valence that aggregates the distribution of valences in $N(i)$ if this distribution satisfies some basic properties, as described bellow. The update is synchronous for all untagged nodes that can be tagged at this time point. The epidemic procedure is repeated until no new valences are assigned to the nodes, i.e. when the population reaches a stable state. The process is similar to those seen in continuous opinion dynamics models~\cite{sirbu2017opinion}, where agents take into account the aggregated opinion of their entire neighbourhood when forming their own. The difference here is that once a valence is assigned, it is never modified.

Let $V$ be the set of all nodes in the network $N$ and $S$ the set of initial seed nodes with their valences $val_s(v_i)$. The aim is to build a set $V^*$ of nodes in $V$ with valences $val(v_i)$ assigned. The procedure is defined in Algorithm \ref{alg:sentimentspreading}.

\begin{algorithm}[t]
\caption{Sentiment Spreading}\label{alg:sentimentspreading}
    \begin{algorithmic}[1]
        \Procedure{sentiment\_spreading}{$V,S,N$}
            \State $V^*=\emptyset$
            \For {$v_i \in S$}\Comment{Initialise valences with seed}
            \State $val(v_i)\gets val_s(v_i)$
            \State $V^* \gets V^* \cup\{v_i\}$
            \EndFor
            \Repeat
            \State $V^*_{old} \gets V^*$
                    \For {$v_i \in V-v^*$}
                    \State $v = \textbf{F}(N(v_i))$ \Comment{$F(N(v_i))$ aggregates the distribution of valences in $N(v_i)$}
                    \If{$v \neq NULL$}
                    \State $val(v_i) \gets v$
                    \State $V^* \gets V^* \cup\{v_i\}$
                    \EndIf
                \EndFor
            \Until{$V^* = V^*_{old} $ }
            \State \bf{return} $V^*$
        \EndProcedure
    \end{algorithmic}
\end{algorithm}

 To decide the aggregation procedure, we took into account several observations. In general, tweets appear to be very heterogeneous, most containing both positive and negative words. Hence a simple averaging of valences would most of the time result in neutral lemmas, although they actually contain meaningful sentiment. So, we decided to use instead the \emph{mode} of the distribution of valences in the neighbourhood, which is a much more meaningful criterion in these conditions. However, the mode was only considered in special circumstances, when the distribution of valences of neighbours was not too heterogeneous. A heterogeneous distribution may have a large range of valences is very large, or the entropy of the distribution can be very high. In this case, it is unclear what the valence of the new lemma should be, so we chose not to assign one at all. Again, this was inspired by works from opinion dynamics (e.g. the q-voter model~\cite{castellano2009nonlinear}), taking into account the concept of social impact: agents are better able to influence their neighbours as a consensual group rather than isolated, hence a heterogeneous group will have no influence on its neighbours. Here, this was implemented as thresholds on the range ($R$) and entropy ($S$) on the neighbouring valence distribution: a node will be infected with the aggregated valence of its neighbourhood only if the range and entropy are bellow these thresholds. To avoid outliers, we consider the range to be the difference between the 10th and the 90th percentile. The two thresholds $R$ and $S$ become two parameters of our model, that need to be tuned to maximise performance.

 Let $range^*$ and $entropy^*$ be the two thresholds. Then the procedure $F$ taking as input $N(v_i)$, the neighbours of $v_i$, can be described by Algorithm \ref{alg:infection}.

\begin{algorithm}[t]
\caption{Infection Function}\label{alg:infection}
    \begin{algorithmic}[1]
        \Procedure{$F$}{$N(v_i),entropy^*,range^*$}
            \State $e=entropy(val(N(v_i)))$ \Comment{entropy of valences of nodes in $N(v_i)$}
            \State $r=range(val(N(v_i)))$ \Comment{range of valences of nodes in $N(v_i)$}
            \If{$e<entropy^* \And r<range^*$}
                \State \bf{return} $mode(val(N(V_i)))$
            \EndIf
            \State \bf{return} NULL
        \EndProcedure
    \end{algorithmic}
\end{algorithm}

With this definition of our spreading model, it is clear that the final valences of the words are influenced by the initial seed lexicon, but also by the structure of the network where emotional valences spread. This structure is determined by the way Twitter users employ the language, hence it depends on the cultural mix in the Twitter community. Thus, the final valences show how the language is used in the corresponding community, from the point of view of the emotional content of words.

\section{Evaluation of emotional valences of words}
\label{sec:evaluation}
To evaluate the emotional valences assigned by EmoSA to different words, we apply the algorithm on tweets in English coming from the UK, from the Untagged Twitter dataset (Section \ref{subsec:untagged_dataset}). The evaluation is based on two criteria. First, we compare the emotional valences assigned by the algorithm with the standard valences in ANEW, by computing the Pearson correlation on the test dataset. Second, we use the community-dependent emotional lexicon for sentiment classification of a Tagged Twitter dataset (Section \ref{subsec:tagged_dataset}) and we evaluate the classification performance. 

It is important to note that the evaluation of emotional valences of words obtained by our algorithm is intrinsically difficult. One hand, we want the valences on the \emph{TEST} dataset to be very close to those in ANEW. On the other hand, however, we expect them to be different, and we actually use the difference as a measure of cultural diversity, since we believe this can be very useful in understanding effects of population migration both on the receiving and incoming population.  We expect that by changing the Twitter population, the correlation between the two changes as well.

\subsection{Evaluation}
\label{sec:word_eval}
\begin{figure}
\centering
\includegraphics[width=0.65\columnwidth]{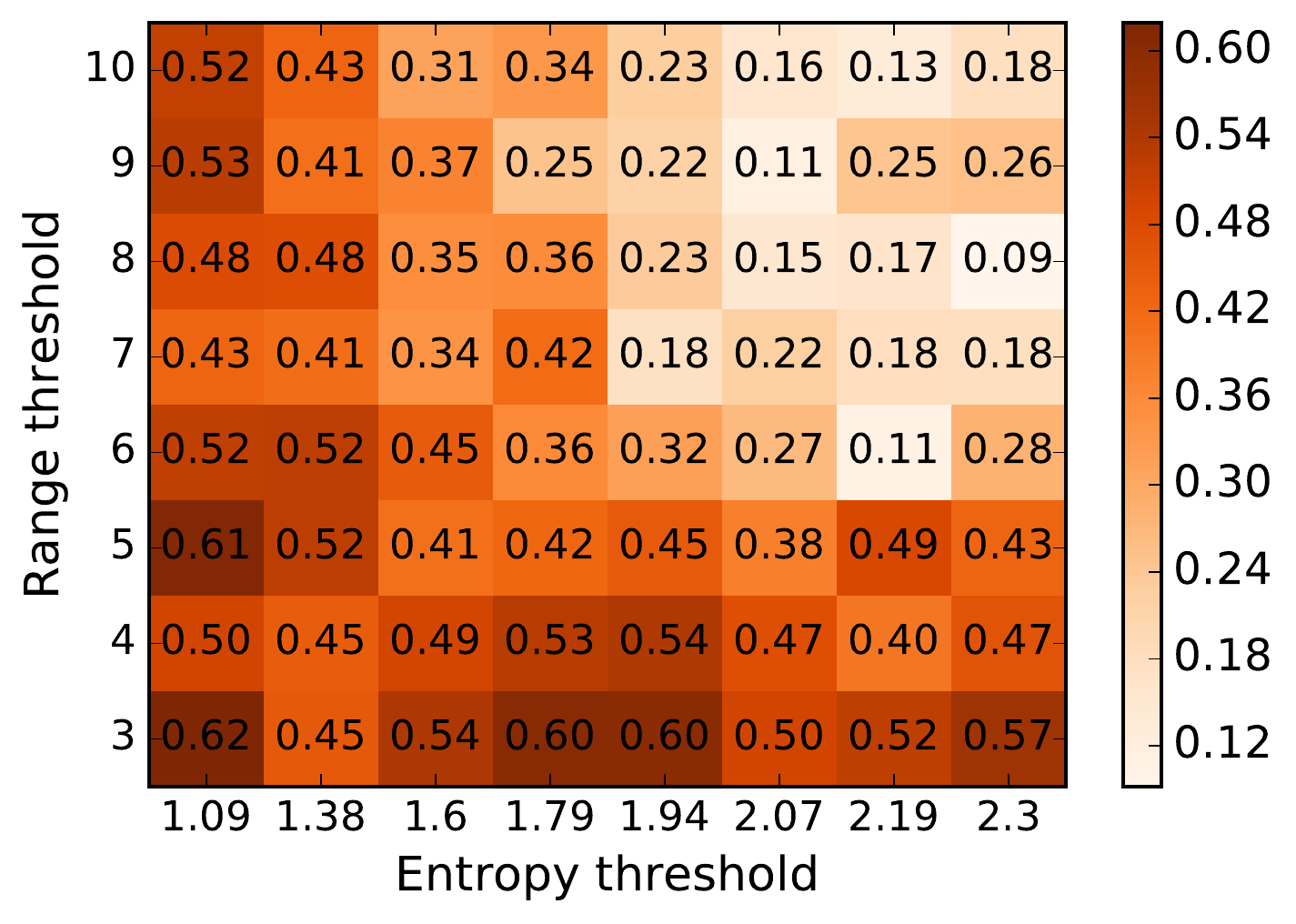}
\caption{Average correlation between modelled and real word valences for the UK data.}
\label{fig:params_UK}
\end{figure}

Figure~\ref{fig:params_UK} shows average correlation values after 10 runs with different test/train splitting of the ANEW dataset, for different values of the thresholds $S$ and $R$. We observe that correlations go up to 0.62, but only when thresholds are very stringent, i.e. only very clear information is allowed to spread in the network. When the distribution of valences in the neighbourhood of a term is too heterogeneous, then it is better not to assign any emotional value to that term. The largest correlation  is obtained for a range threshold of 3, and entropy threshold of 1.09 (we consider the distribution described by 10 bins of equal size, hence the maximum entropy is approximately 2.3), which are the parameters used in the rest of the paper for UK data.

\begin{figure}[t]
\centering
\includegraphics[width=0.55\textwidth]{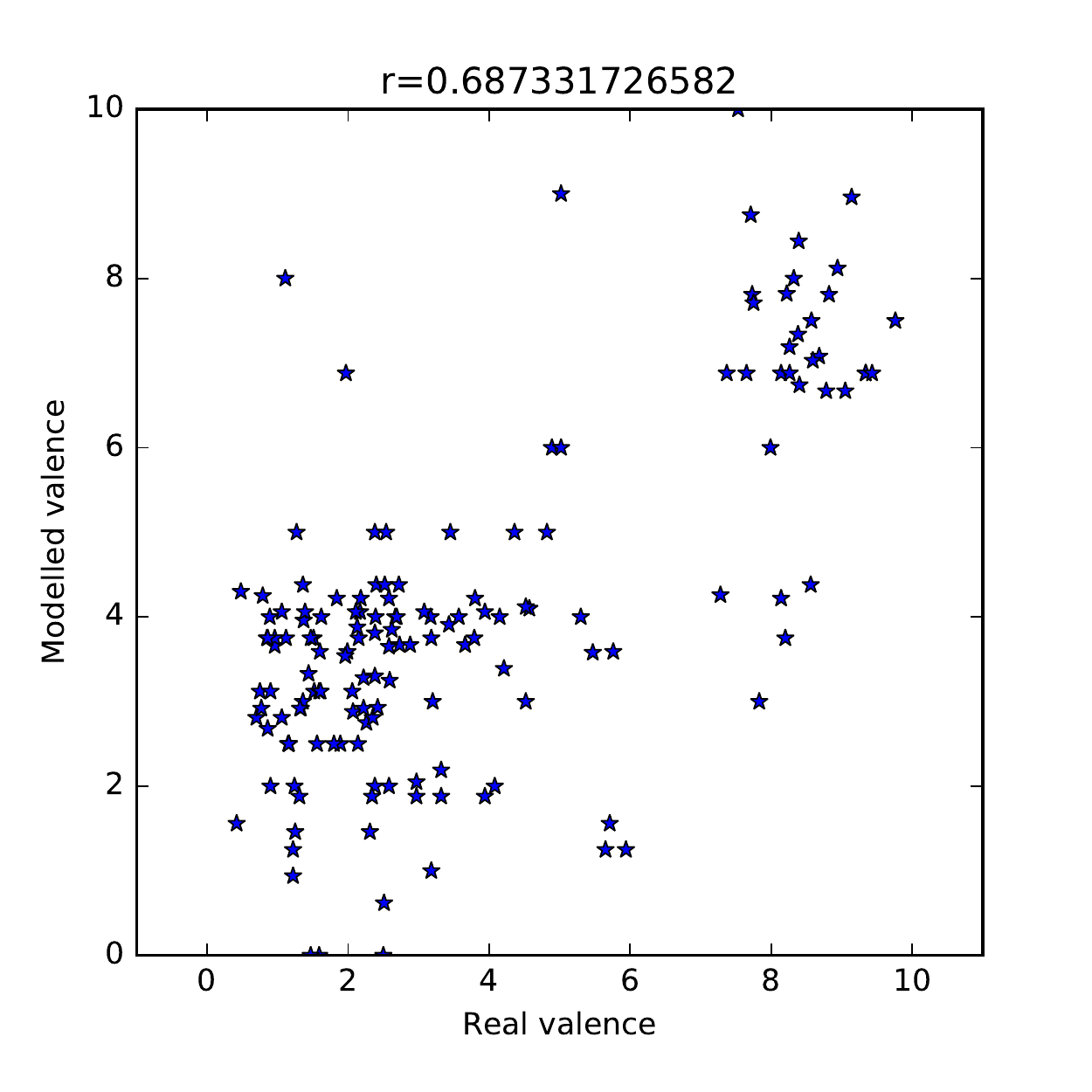} 
\caption{Modelled and real word valence for one selected run with best parameters.}
\label{fig_corr}
\end{figure}

\begin{figure}
  \centering
\includegraphics[width=.40\textwidth]{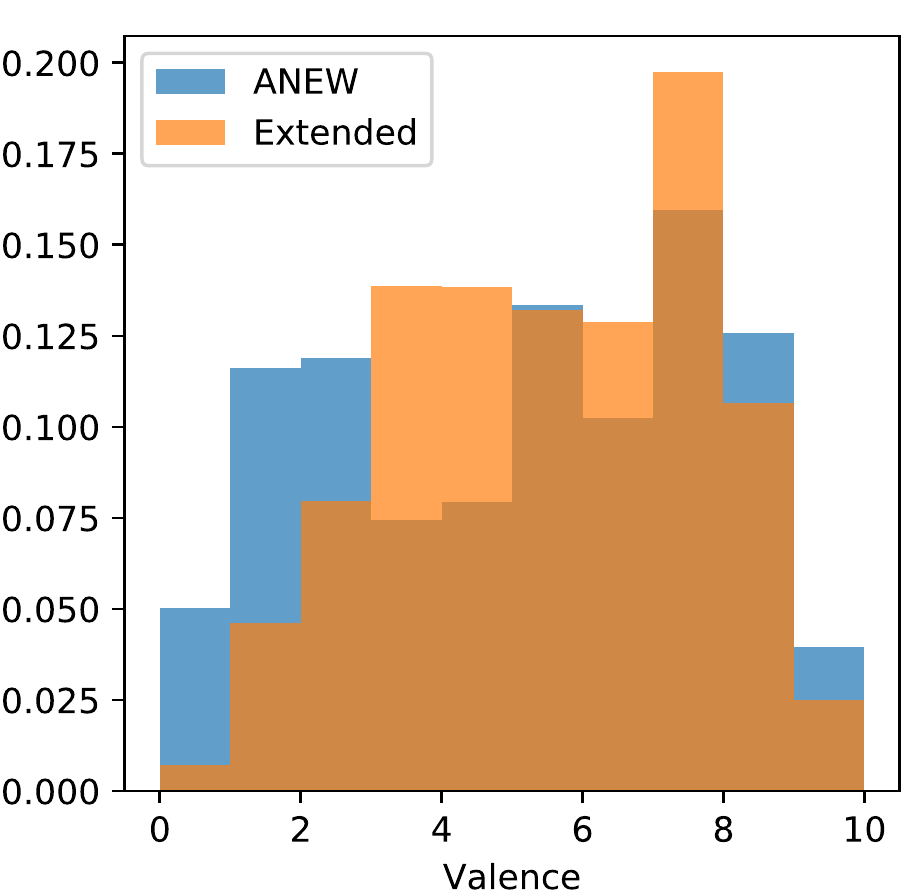} 
\caption{Histogram of valences for ANEW and CD EmoLex.}
\label{fig_histdict}
\end{figure}

Figure~\ref{fig_corr} displays the modelled and real valences on test data for the run with the best correlation. The plot shows clearly that the valences obtained by our method align well with human-tagged data, validating our approach.  For further comparison, we also display the distribution of valences in the ANEW lexicon, compared to the community-dependent emotional lexicon obtained through our method (Figure~\ref{fig_histdict}). We observe some differences here. While negative lemmas make a small fraction of the lexicon in both cases, the CD EmoLex still contains a larger fraction of neutral and positive lemmas, compared to the ANEW lexicon. This is, however, not a concern, given that we observed this trend for other lexicons as well (see description of SentiWordNet in Section~\ref{sec:lexical_res}).

A second criterion for validation of the CD EmoLex is classification performance on the Tagged Twitter Dataset (Section~\ref{subsec:tagged_dataset}). We implemented a sentiment classifier based on Support Vector Machines (SVM), that used several features to classify sentiment of tweets into three classes: negative, neutral and positive. 
The features used include, for each tweet, several statistics over the valence of individual lemmas contained by the tweet: arithmetic and geometric mean, median, standard deviation, minimum and maximum. To these, we added the number of lemmas with a valence over 7 and over 9, to understand the presence of positive terms. Conversely, we also computed the number of lemmas with a valence under 3 and under 1. Finally, we included the total length of the tweet, and a boolean feature flagging the presence of a negation. We only considered tweets for which at least 3 lemmas were found in the lexicon.

\begin{figure}[t]
\centering
\includegraphics[width=1\textwidth]{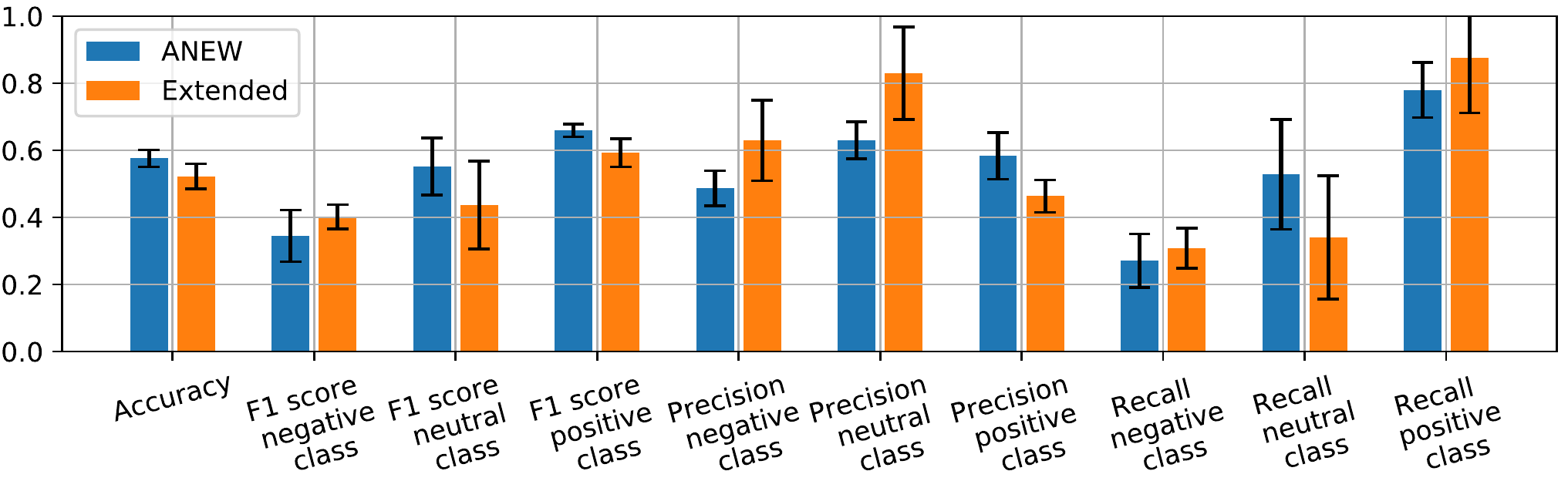} 
\caption{Performance of SVM classifier using the original ANEW lexicon only and our community-dependent emotional lexicon.}
\label{fig_svm}
\end{figure}

 The features above can be computed using any lexicon, and SVM performance can vary when changing the lexicon. We compare the performance of our CD EmoLex (described in Figures~\ref{fig_corr} and \ref{fig_histdict}) with that of ANEW. We expect no decrease in performance with our community-dependent emotional lexicon. To validate results, we used a cross-validation approach, where 80\% of tagged tweets were used to train the SVM and 20\% to test it. The analysis was repeated ten times for each lexicon, with average performance displayed in Figure~\ref{fig_svm}. Error bars show one standard deviation from the mean. The plot shows that the performance with the CD EmoLex is comparable to ANEW, validating our approach. The F1 score increases on the negative class and decreases on the other two. Precision increases on negative and neutral, while recall increases on negative and positive tweets. Hence, our community-dependent emotional lexicon seems to perform slightly better on negative tweets, however it overestimates the positives. Given that negative tweets are a small part of the corpus, accuracy decreases slightly. The performance range across repeated runs always overlaps between the two lexicons compared.
 
\section{Superdiversity Index: computation and evaluation}
\label{sec:SI_eval}

Since we observed in the previous Section that the emotional valences assigned to terms by our algorithm are in agreement with the standard valence, we now proceed to check whether differences between communities can be related to cultural diversity in a population. We compute the $SI$ at different space resolutions for the UK and Italy, and consider as a baseline the foreign immigration rates in the same geographical regions as those where the $SI$ was computed. The immigration rates were extracted from the D4I dataset (Section~\ref{sec:d4i}). We expect that communities with higher immigration rates also have higher diversity so higher $SI$. We calculate the Pearson correlation ($r$) between the $SI$ values and immigration rates, as a measure of the performance of the proposed $SI$.

\subsection{Computing the SI}
Algorithm~\ref{alg:si} shows the steps needed to calculate the $SI$. After applying EmoSA $compute\_valences$, for each term $t$ in the $TEST$ dataset (with valence $v_t$) we have a corresponding modelled valence $v^*_t \in TEST^*$, which tells us how the word is actually used by the community. We then compute the Pearson correlation, $r = pearson (TEST, TEST^*)$, between the standard and the community-dependent valences of words over all the words in the $TEST$ dataset.

\begin{algorithm}[t]
\caption{SI computation}\label{alg:si}
    \begin{algorithmic}[1]
        \Procedure{superdiversity\_index}{$ANEW,UTD,iteration\_count$}
        \State $\bar{r} \gets 0$ \Comment {Average Pearson correlation}
        \For {$i \in 0..iteration\_count$}\Comment{Repeat many times}
            \State $TRAIN,TEST \gets split(ANEW,0.5)$ \Comment{Split ANEW 50\%-50\%}
            \State $TEST^* \gets compute\_valences(TRAIN,TEST,UTD)$ \Comment{Compute valences from Twitter data}
            \State $r\gets pearson(TEST^*,TEST)$ \Comment{Compute correlation between modelled and real values}
            \State $\bar{r}\gets \bar{r}+ r$
            \EndFor
            \State $\bar{r}\gets \bar{r}/iteration\_count$
            \State $SI \gets (1-\bar{r})/2 $ \Comment{Compute $SI$ from correlation}
             \State \bf{return} $SI$
        \EndProcedure
    \end{algorithmic}
\end{algorithm}

In this work, we use ($iteration\_count = 10$). $SI$ takes values in the range $[0,1]$.

\subsection{SI evaluation}
To prove that it is the community that determines the value of the SI, and that our correlations are not random, we also devise a \emph{null model SI}. This is achieved by reshuffling tweets across geographical regions, maintaining fixed the number of tweets in each region. We compute the correlation among the null model $SI$ and immigration rates, and compare them with the original $SI$ correlation.

An additional evaluation step compares the performance of the newly proposed $SI$ with that of other possible superdiversity measures extracted from the same data. We consider five additional measures. The first two relate to the frequency of tweeting. One could hypothesise that a more diverse community would tweet more, or less. Hence we consider the total number of tweets in the local language, together with the population-normalised version, i.e. number of tweets per capita. A second category of measures relates to the different languages spoken by a Twitter community. We would expect that a more diverse community will use more languages. We consider the absolute number of languages, but also the entropy of the distribution of tweets in the various languages. The latter measure takes into account the volume of tweets in each language, besides the number of languages. The fifth possible measure of diversity relates to the lexical richness of the language used by a twitter community. Again, one could expect a richer language from a diverse community. To quantify this, we use a well-known index used in Linguistics, the Token Type Ratio (TTR)~\cite{templin1957certain}. It is computed as the ratio between the number of token types (in our case the number of different words) that make up the vocabulary and the overall size of the corpus (the total number of words in the corpus). The TTR value ranges in [0,1], where values closer to 1 denote that texts are varied and rich.

\subsection{UK}

\begin{figure*}
\includegraphics[width=0.5\columnwidth]{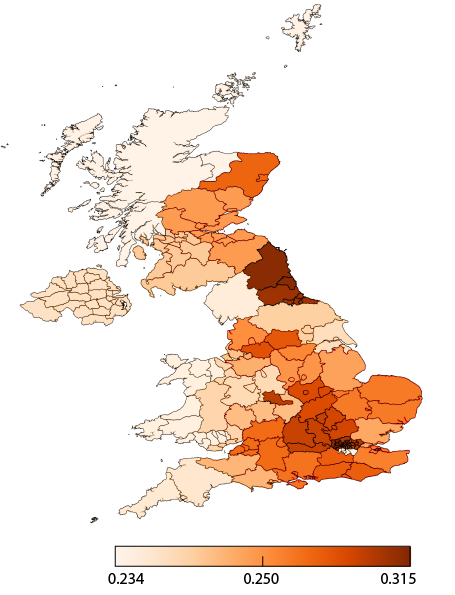}
\includegraphics[width=0.5\columnwidth]{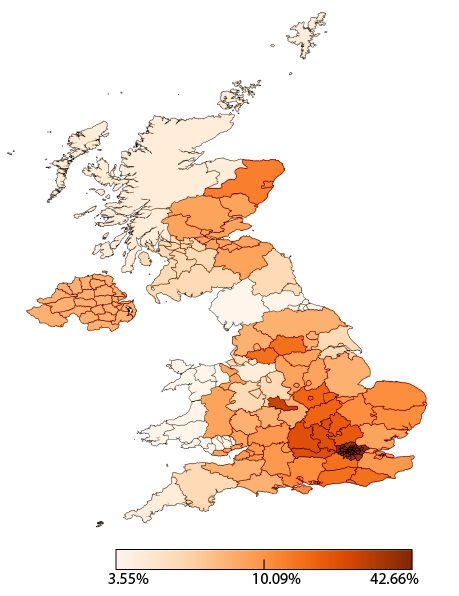}
\caption{Superdiversity index (left) and immigration levels (right) across UK regions at NUTS2 level.}
\label{fig:map_UK}
\end{figure*}

\begin{figure*}
\includegraphics[width=0.3\textwidth]{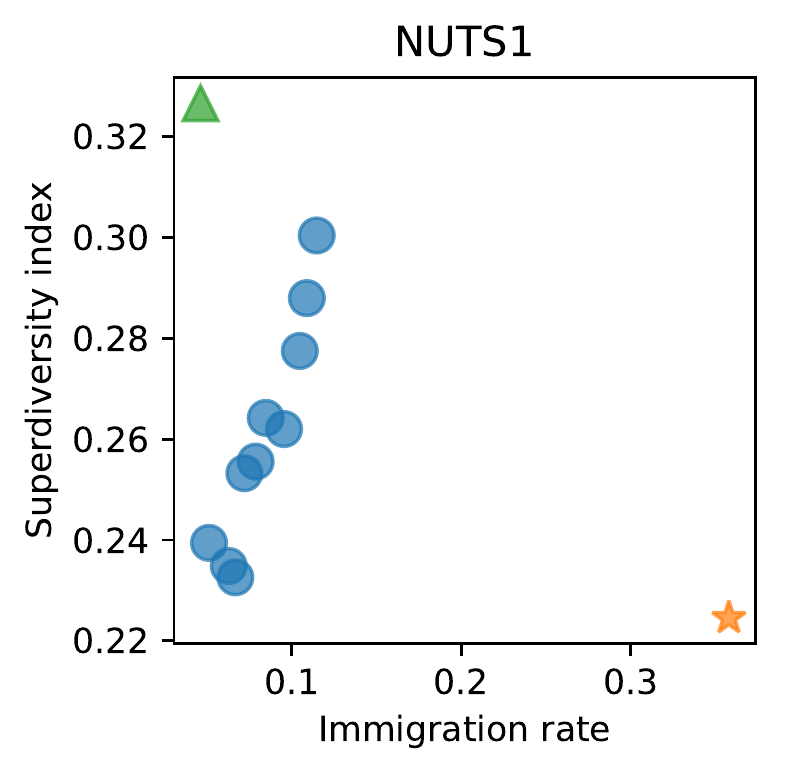}
\includegraphics[width=0.3\textwidth]{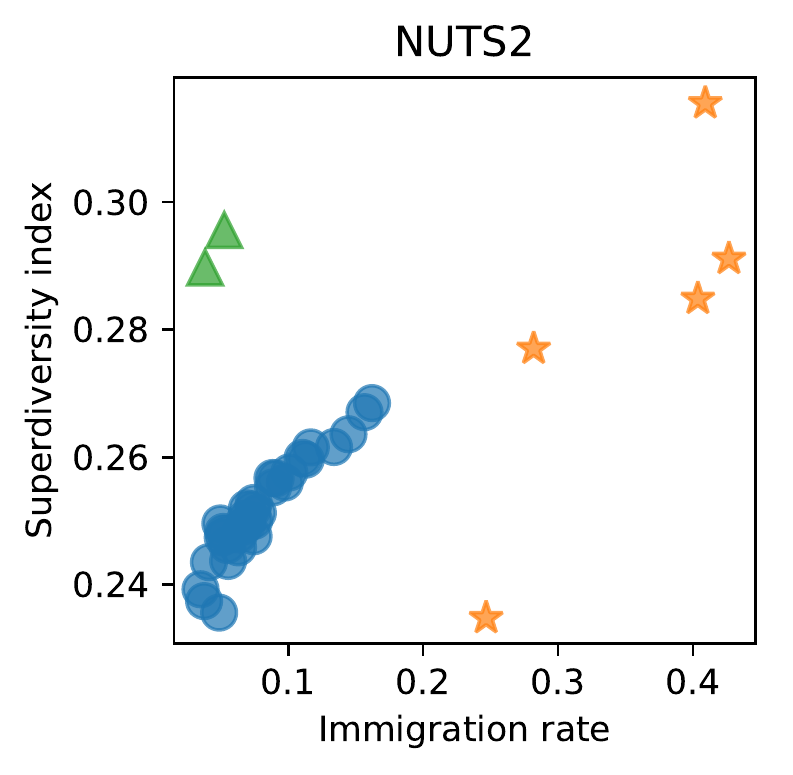}
\includegraphics[width=0.3\textwidth]{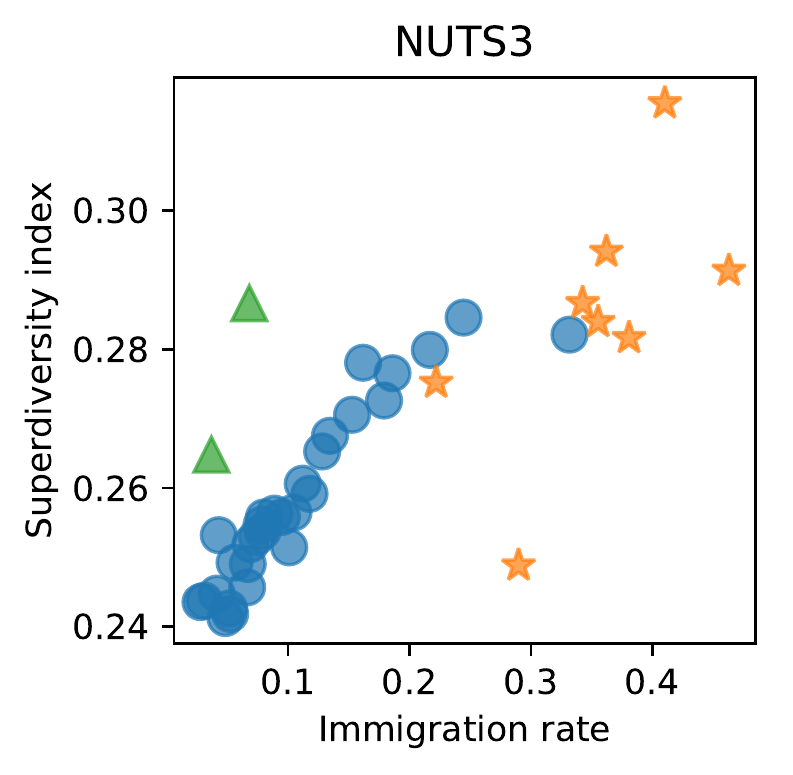}
\caption{SI values versus immigration rates at different geographical levels, for the UK. At the level NUTS3 we selected the top 40 regions based on the number of tweets available in the dataset. The stars correspond to the London area, the triangles represent regions in Northeast England, while the rest of the regions are displayed with circles. }
\label{fig:GBR}
\end{figure*}

The proposed $SI$ was computed for the UK at three different geographical resolutions. The NUTS1 level corresponds to 12 UK regions while the NUTS2 level to 40 regions. The NUTS3 level contains 174 different regions, out of which we select the 40 with the largest number of tweets. For computation of the $SI$ we considered all the tweets in English published in the various regions. Figure~\ref{fig:map_UK} shows visually the geographical distribution of $SI$ values at level NUTS2, and compares with the distribution of foreign immigration levels, from the D4I dataset. There is a clear similarity among the two maps. To understand better the relation between $SI$ and immigration rates, Figure~\ref{fig:GBR} plots the $SI$ values obtained versus the immigration rates, at each NUTS level analysed. We observe that most of the regions align very well on a line, with a very large correlation with the immigration rates. 

At all geographical levels, we observe that the regions corresponding to Northeast England and the London area appear to have a different behaviour, deviating from the main line defined by the other regions. This is also visible on the map, at regional level. However, when moving from NUTS1 to NUTS2 and NUTS3, we see that, within the two regions, $SI$ grows as the immigration rate grows. For instance, at level NUTS2, when considering only the five regions from the London area, we see that  $SI$ is larger when immigration is larger. The same is true for the regions from Northeast England. We believe this is due to different ranges of $SI$ in different areas, and will discuss it further in Section~\ref{sec:discussion}. For the rest of this section, we will consider only the remaining regions, i.e. all UK except for Northeast England and the London area. Table~\ref{tab:UK} shows the exact values of the Pearson correlation between immigration rates and the SI, which prove to be remarkably well correlated at all geographical levels.  The null model $SI$ does not correlate at all, as expected, giving evidence that the correlations we obtained are meaningful and related to the source community of the tweets, and not merely due to the number of tweets in each region.

The comparison with other possible measures of diversity is also very favourable to our proposed SI, which is clearly superior to all others, as the same Table~\ref{tab:UK} shows. No relation between immigration rates and frequency of tweets appears to exist. Some correlation seems to emerge with the number of tweets per capita, number of languages and the language entropy, however much lower than  the $SI$ case, and not stable at all geographical levels. 

\begin{table*}
	\centering
    \begin{tabular}{|p{1.5cm}|p{0.8cm}|p{0.8cm}|p{1cm}|p{1.4cm}|p{1cm}|p{1cm}|p{0.8cm}|}
	\hline
	Geographical level  & $SI$ & null model $SI$ & number of English tweets &number of English tweets per capita& number of languages & language entropy & TTR \\
	\hline 	\hline
    NUTS1 (10 regions) & \bf{0.943} &-0.236& 0.328& -0.520& 0.519& 0.481& -0.005 \\ 
	\hline
 NUTS2 (40 regions) & \bf{0.941}&-0.137& 0.332& 0.007& 0.362& 0.288& -0.340 \\
 	\hline
 NUTS3 (40 regions)& \bf{0.928}& -0.221& 0.141& 0.049& 0.322& 0.529& 0.147 \\
	\hline
	\end{tabular}
	\caption{Correlation between different measures of diversity extracted from Twitter and the immigration rates, at various geographical levels in the UK, excluding London and Northeast England. At the level NUTS3 we selected the top 40 regions based on the number of tweets available in the dataset. }
	\label{tab:UK}
\end{table*}

\subsection{Italy}

\begin{figure}
\centering
\includegraphics[width=.65\textwidth]{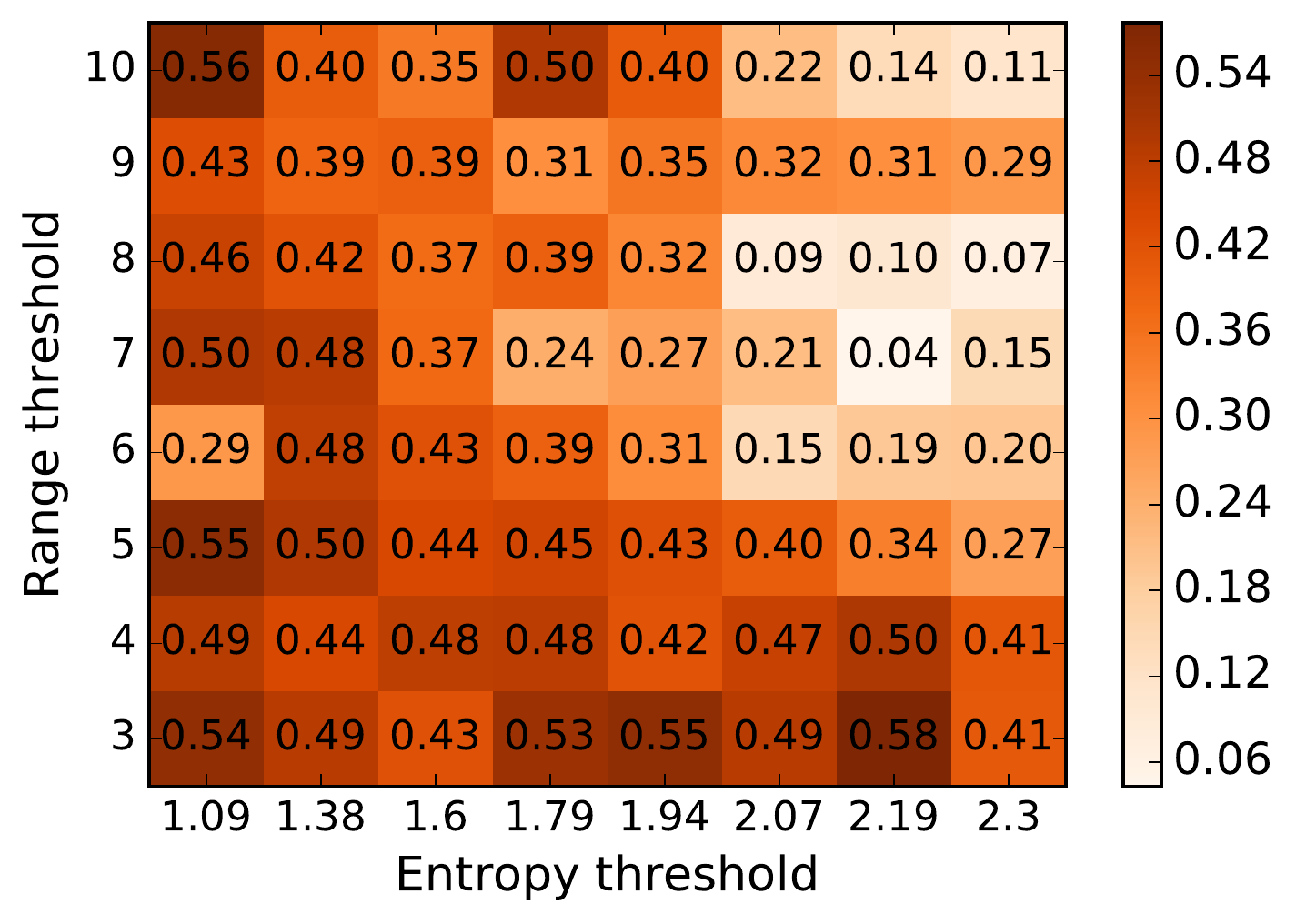}
\caption{Optimisation of model parameters for Italy.}
\label{fig:params_ITA}
\end{figure}

The analysis above was repeated for Italy. We translated the tagged dictionaries into Italian and applied the method to all Italian tweets published from different Italian regions. We first report results for various $S$ and $R$ parameters.  Figure~\ref{fig:params_ITA} shows that best results were obtained for $R=3$ and $S=2.19$, which are the values used to obtain all the results related to Italy presented in this section.

\begin{figure*}
\includegraphics[width=0.5\columnwidth]{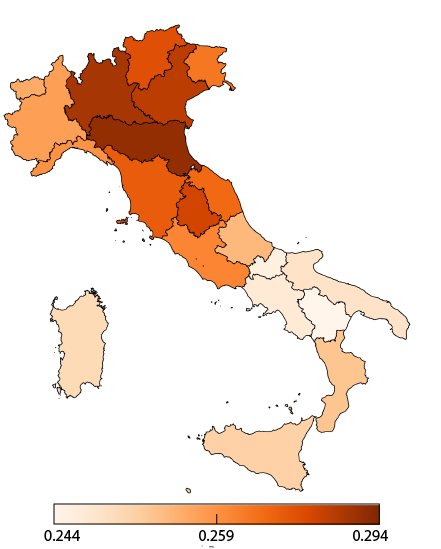}
\includegraphics[width=0.5\columnwidth]{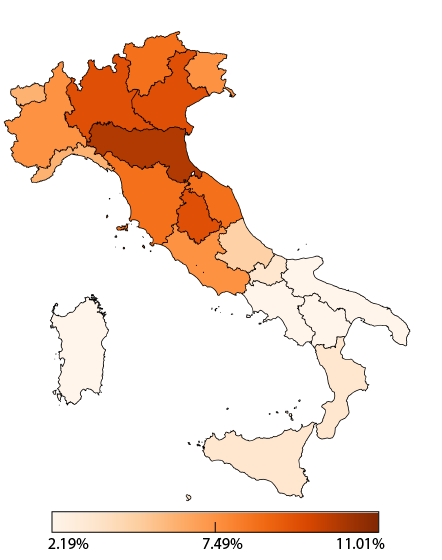}
\caption{Superdiversity index (left) and immigration levels (right) across Italian regions at NUTS2 level.}
\label{fig:map_ITA}
\end{figure*}

\begin{figure*}
\includegraphics[width=0.3\textwidth]{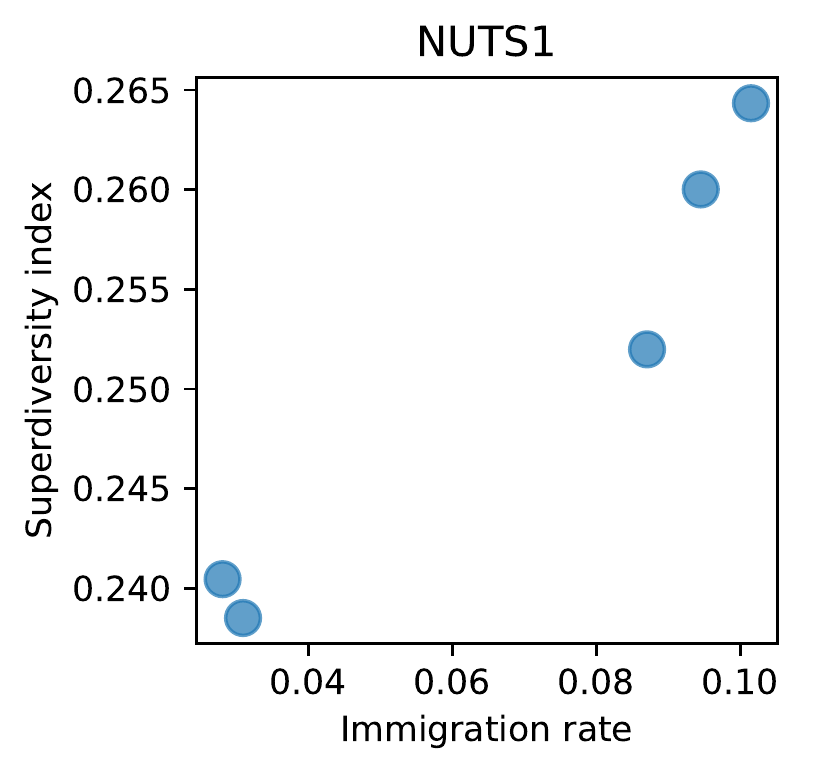}
\includegraphics[width=0.3\textwidth]{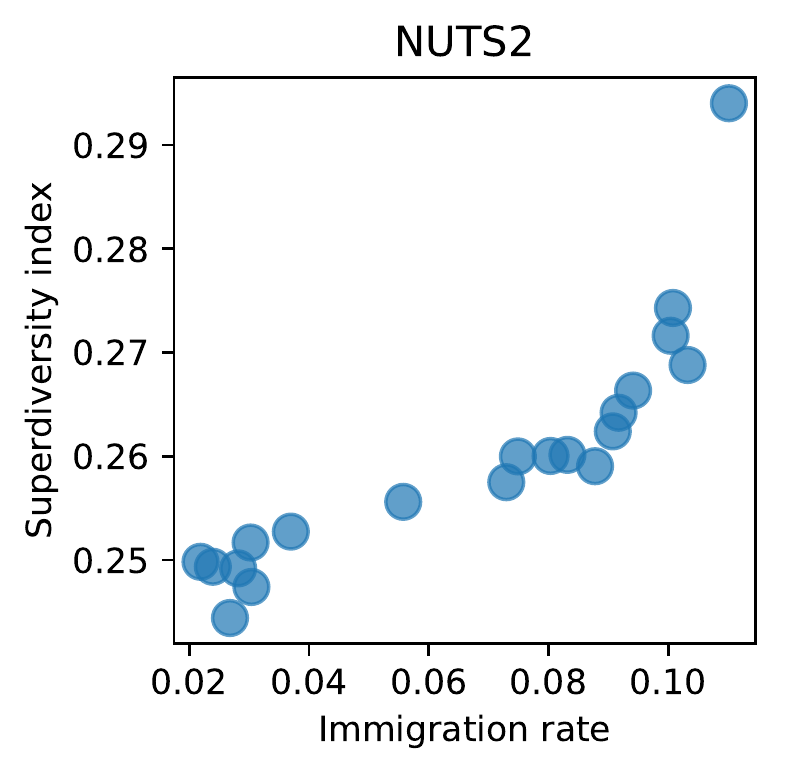}
\includegraphics[width=0.3\textwidth]{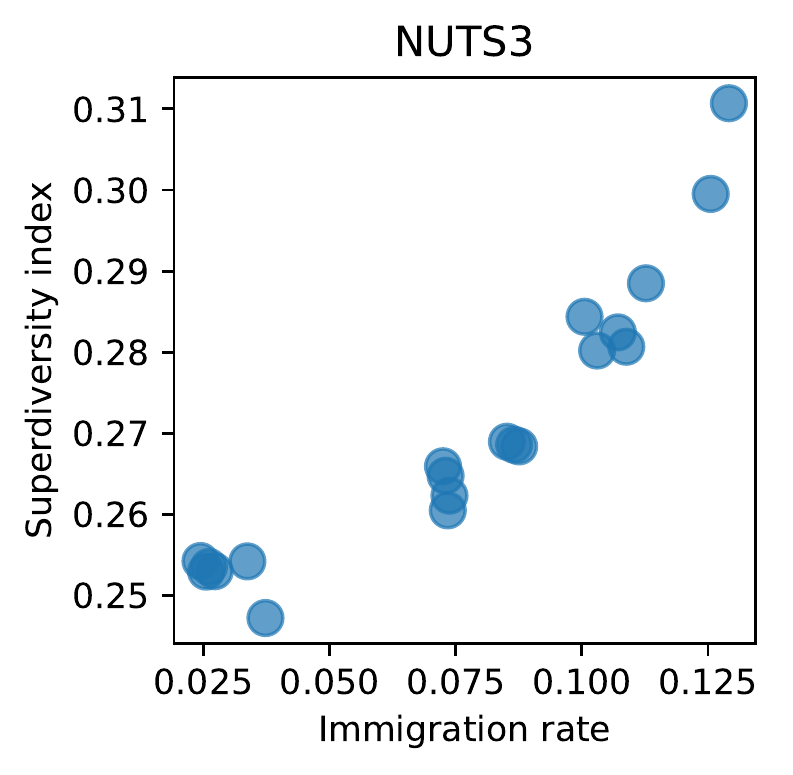}
\caption{SI values versus immigration rates at different geographical levels, for Italy. At the level NUTS3 we selected the top 20 regions based on the number of tweets available in the dataset.}
\label{fig:ITA}
\end{figure*}

\begin{table*}
	\centering
    \begin{tabular}{|p{1.5cm}|p{0.8cm}|p{0.8cm}|p{1cm}|p{1.4cm}|p{1cm}|p{1cm}|p{0.8cm}|}
	\hline
	Geographical level  & $SI$ & null model $SI$ & number of Italian tweets &number of Italian tweets per capita& number of languages & language entropy & TTR \\
	\hline
	\hline
    NUTS1 (5 regions) & \bf{0.963}& -0.437& 0.735& 0.696& 0.183& -0.585& -0.727 \\ 
	\hline
 NUTS2 (20 regions) & \bf{0.859}& 0.143& 0.279& 0.282& 0.304& 0.099& -0.243 \\
	\hline
  NUTS3 (20 regions)& \bf{0.924}& 0.082& 0.081& -0.148& 0.216& 0.021& 0.091 \\
	\hline
	\end{tabular}
	\caption{Correlation between different measures of diversity extracted from Twitter and the immigration rates, at various geographical levels in Italy. At county level (NUTS3) we selected the top 20 regions based on the number of tweets available in the dataset. }
	\label{tab:ITA}
\end{table*}

Figure~\ref{fig:map_ITA} shows the geographical distribution of both the $SI$ values and the immigration rates from the D4I dataset, at level NUTS2 (regional). Again, there is a very good similarity among the two maps. In Figure~\ref{fig:ITA} we plot the $SI$ values obtained by our method versus the immigration rates. Excellent correlation with immigration can be observed, at all geographical levels. Exact correlations are reported in Table~\ref{tab:ITA}, with values over 0.85 at all levels.  Please note that at level NUTS1, Italy is divided into only 5 regions, hence here correlations are not really meaningful. We report them for completeness, however is is the behaviour at the following levels that carries significance.  At NUTS2 there are 20 regions, while at NUTS3 we select the top 20, similar to the UK case.

The table also shows results for the null model, which are as expected: the null model $SI$ does not correlate with immigration rates. As for the other possible diversity measures, none of them seem to give any hint of the immigration rate, at levels NUTS2 and NUTS3. Hence our proposed $SI$ is undoubtedly superior. At level NUTS1 we see some correlation, but again we believe these values to be spurious, since we are considering only 5 geographical areas.

\section{Discussion}
\label{sec:discussion}
The results presented above are very promising, and show a strong link between the proposed $SI$ and foreign immigration rates. However, in the UK case, a small number of regions seemed not to show the same behaviour as the rest of the country. These were the London Area and Northeast England. In the first case, although immigration rates in London are much higher than the rest of the country, the $SI$ value extracted from all tweets in London was smaller (NUTS1). However, when dividing tweets per county, within the London area, $SI$ values seem to grow as immigration levels grow (NUTS2 and NUTS3). Hence, it appears that the ranges of the $SI$ obtained are different in London compared to the rest of the country. The same applies to Northeast England. Even if immigration rates are low, $SI$ values were high. Again, those regions seem to form a cluster of their own, where $SI$ ranges do not match those from the rest of the UK. Thus, in order to make the $SI$ range uniform, we need to identify correcting factors or at least determine the various clusters, without using the immigration rate itself. This should also make $SI$ values comparable across geographical resolutions, since we observe that, for the same immigration rate, $SI$ values can vary from one NUTS level to another.  

\begin{figure}
\centering
\includegraphics[width=0.6\columnwidth]{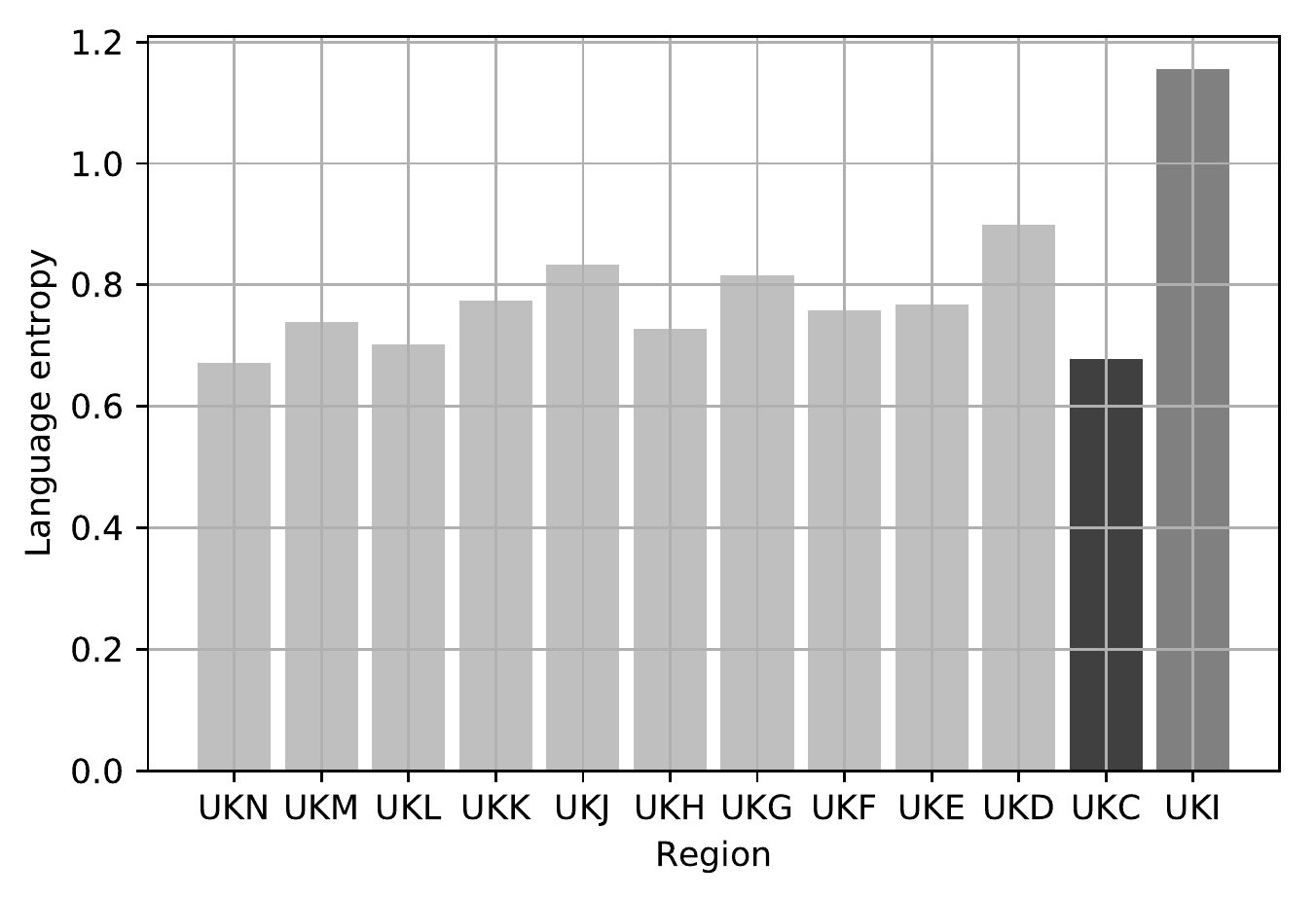}
\caption{Language entropy on tweets originating from the UK, at macro scale (NUTS1 regions). The region UKI corresponds to the London area, while UKC to Northeast England.}
\label{fig:entropy_UK}
\end{figure}

\begin{figure}
\centering
\includegraphics[width=0.6\columnwidth]{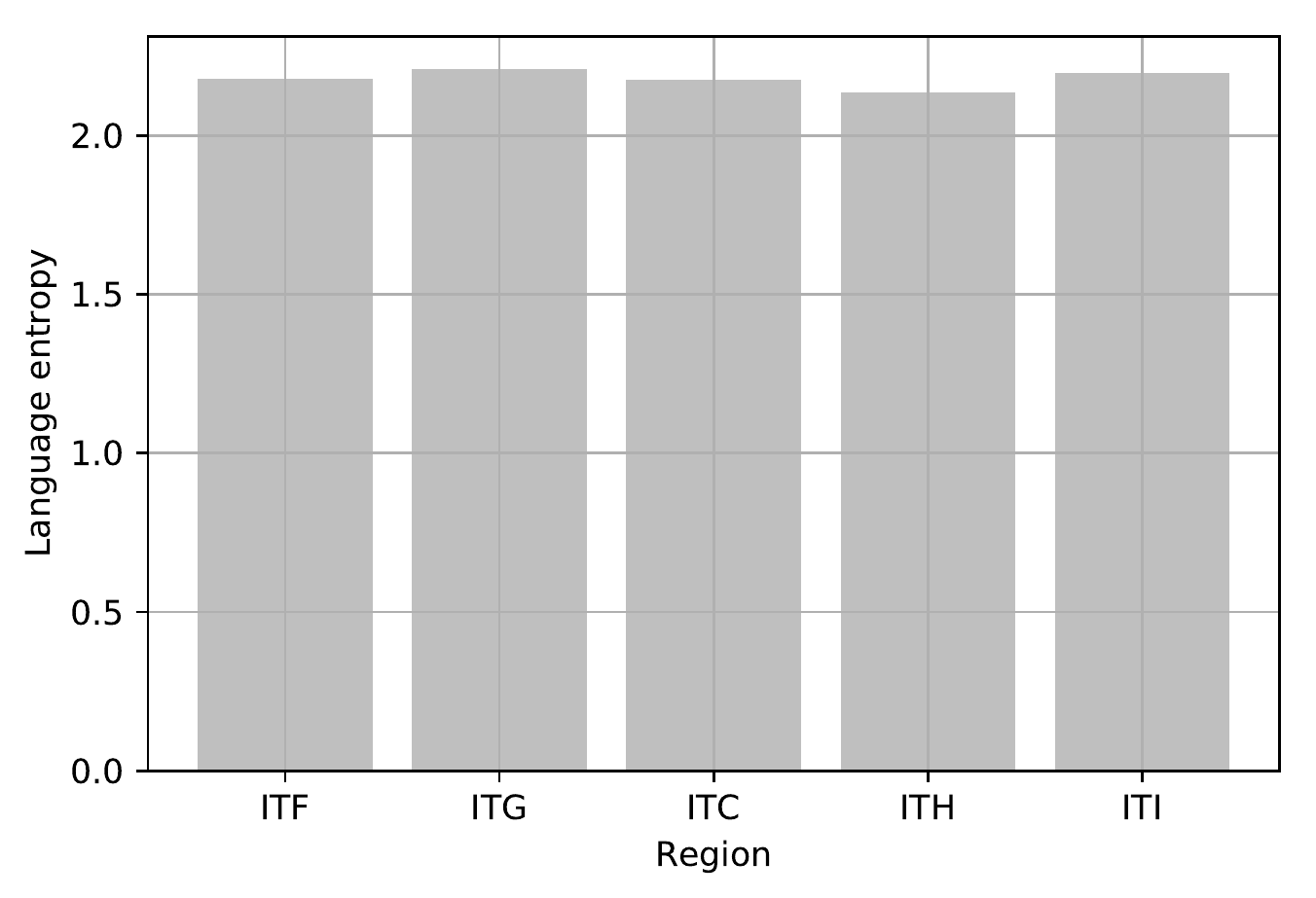}
\caption{Language entropy on tweets originating from Italy, at macro scale (NUTS1 regions). }
\label{fig:entropy_ITA}
\end{figure}

One possible correcting factor could be language entropy. We observed that the London area displays, at NUTS1 level, a much higher language entropy compared to the rest of the country, hence it can be used to rescale the SI. Figure~\ref{fig:entropy_UK} shows language entropy for all NUTS1 regions. In Italy, however, language entropy is very similar across regions (figure~\ref{fig:entropy_ITA}), and in fact there seem not to be any range issues here, since all regions overlap very well. The uniform and large entropy in Italy could be explained by the fact that Italy is a popular tourist destination, so many tweets in different languages exist. For instance in central Italy we have 25044 tweets in English and 67903 in Italian, a factor of only 2.7 between the local and a foreign language.  

A different important factor could be cultural differences of the local non-immigrant community, such as those identifiable by local dialects. A higher use of the local dialect could explain the apparent larger  $SI$ values in Northeast England. To correct for this, a baseline $SI$ value could be computed from a subset of tweets coming from local users (for example local newspapers or official accounts). This baseline could be used to correct for range differences to the rest of the country.  

Given the high correlation to foreign immigration rates, we believe that our $SI$ is a first step towards a novel nowcasting model for migration stocks. Once all correcting factors are identified, these could be used, together with the SI, to build a highly accurate model to estimate immigration rates. We expect that a standard machine learning model could prove suitable for this task, that we plan to undertake in future work. Such a model will enable accurate immigration statistics without the need for time- and resource-consuming population censuses. By repeating the analysis regularly, we will be able to maintain updated statistics, valid also in regions where censuses are not possible or inaccurate due to clandestine immigration.

\section{Related Work}
\label{sec:related}

The migratory phenomenon intrinsically concerns at least two countries; therefore, its study implies the use of cross-referenced data coming from among various nations. To date,  most of the information about migration flows and stocks derive from administrative data and official statistics, thus from surveys and national censuses. Databases coming from different countries are often inconsistent among themselves and show low spatial coverage and reduced time resolution. 
The recent wide availability of Big Data allowed to employ these new data in understanding and estimating the migration phenomenon and to observe related consequences with high resolution and spatial coverage. Thanks to their characteristics, it may be possible to extract new indexes for estimating, nowcasting, and evaluating stocks and flows and integration of immigrants.
These data can highlight most of the important effects of migration on both the migrant population and on the receiving and source communities. 

In Human Migration studies, works exploiting Big Data have been carried out using, among others, mobile phone data,  e-mail communication logs, Skype data, and social media data. Notably, the latter contains various types of information about users. As compared to traditional data sources, they are potentially unlimited in terms of sample size since they cover a broad set of the worldwide population. Online Social Networks (OSNs) can allow researchers to analyze temporal variations easily, and to now-cast migration phenomena. 

The literature on social media data for Human Migration studies includes multiple social platforms. Rodriguez et al. \cite{rodriguez2014migration} and Barslund et al. \cite{barslund2016mobile} used data gathered from LinkedIn to identify mobility patterns of highly skilled migrants. Messias et al. \cite{Messias2016} employed Google+ data to study location patterns of migrants who have lived sequentially in more than two countries. Finally, in less recent works, State et al. \cite{state2013studying} and Zagheni et al. \cite{zagheni2012you} exploited Yahoo!'s users' data to infer global flows of both migrants and tourists. 

However, most studies focused on data from Twitter and, more recently, Facebook. Herdagdelen et al. \cite{HerdagdelenAmac2016} inspected friendship ties of immigrants in the United States (US) to study the composition of their social networks on Facebook. Dubois et al. \cite{Dubois2018} and Zagheni et al. \cite{zagheni2017leveraging} used data from Facebook's advertising platform to estimate assimilation of Arabic-speaking migrants in Germany, and stock of international migrants in the US, respectively. Stewart et al. \cite{stewart2019rock} analyzed musical interests on Facebook to estimate the cultural assimilation of Mexican immigrants in the US. Finally, Spyratos et al. \cite{spyratos2018migration} computed independent estimates of Facebook Network ``expats'' by correcting the selection bias of the platform users compared to the real population.

Regarding Twitter, Steiger et al. provided a survey of using Twitter data for research \cite{steiger2015advanced}, reviewing 92 research papers. The study shows that 57\% of the papers take into account only semantic information (e.g., hashtags, followers, and profiles), that 33\% rely on both semantic and spatio-temporal information, and that the 10\% exploits only spatio-temporal information. Related to migration, Hawelka et al. \cite{hawelka2014geo} and Zagheni et al. \cite{zagheni2014inferring} exploited geo-localized tweets to analyze trends in both mobility and migration flows. Fiorio et al. \cite{fiorio2017using} estimated US internal migration flows in different timing and durations and showed, for the first time, relationships between short-term mobility and long-term migration. Freire-Vidal et al. \cite{freire2019characterization} characterized the behavior of locals of Chile towards immigrants and measured differences between attitudes by using psycho-linguistic lexicons and interaction networks. 

Although the literature has highlighted potential ethics, selection bias, and privacy issues, various research attested the importance of social media data sources in understanding migration patterns. 
For instance, in a recent work, Aswad et al. \cite{aswad2018refugee} investigate the role of social media to understand whether it can be considered a proxy for immigration. The data analysis confirms that Twitter may be regarded as a valid source for information about immigration and refugee placement.

The data coming from OSNs can enable the study of the migratory phenomenon also through the language used by users \cite{lamanna2018immigrant,hubl2017analyzing}. Indeed, the language could represent a direct link between individuals, origin country, and nationality. Moreover, the language allows observing how linguistic characteristics of people vary when in contact with other cultures. 
The analysis of the language used in social networks applied to human migration studies involves several research fields, including but not limited to mobility patterns \cite{mocanu2013twitter}, migration stocks and flows \cite{magdy2014exploiting,lamanna2018immigrant}, and sentiment analysis \cite{senaratne2014moving}.

Mocanu et al. \cite{mocanu2013twitter} proposed a global analysis of indicators and linguistic trends starting from Twitter. By aggregating data based on different geographical scales, the study characterized the worldwide linguistic geography finding a universal pattern describing users' activity across countries. The authors studied various indicators, such as the linguistic homogeneity of the different countries, the seasonal tourist patterns within the countries, and the geographical distribution of the different languages in the multilingual regions. The proposed analysis suggests that there would be a high correlation between the heterogeneity of Twitter penetration and the Gross Domestic Product (GDP). The results also show that the statistical usage model of the social platform is independent of factors such as country and language. Finally, by analyzing the temporal variations of the linguistic composition by country, it is possible to observe travel patterns and identify seasonal and real-time mobility patterns.

Magdy et al. \cite{magdy2014exploiting} proposed an analytical study of relations between the language of tweets and their spatial distribution to investigate a) language diversity and its usage in Twitter communities geolocalized by country; and b) spatial distribution of cultural groups in different countries. Starting from the identification of local communities, the authors identified dominant languages and the spatial distribution of minor languages in local communities. Linguistic diversity is calculated for each identified local community through the use of statistical measures. Findings show that measures taking into account the language distribution within the community are more coherent and robust in identifying the  countries with the highest language diversity. Furthermore, by comparing linguistic diversity measures with official data, the authors showed that Twitter's community sample could be representative of the actual population. Finally, based on the spatial distribution within the countries, the study identified different cultural groups. This task could be useful to understand certain behaviors related to specific cultural groups such as Syrian refugees preferring to get closer to communities with similar cultural traits.

Gon\c{c}alves et al. \cite{gonccalves2014crowdsourcing} exploited geolocalized tweets to study diatopic variations of the modern-day Spanish language. The authors applied machine learning techniques to inspect a large dataset of Spanish vulgar tweets. Through these analyses, it is possible to identify, for the first time, two varieties of the Spanish language. It seems that the Spanish language attested on Twitter has two super dialects: an urban speech typically used in the main Spanish and American cities, and a local speech mostly attested in small towns and rural areas. The proposed approach also allowed the authors to identify regional dialects and their approximate isoglosses.

Moise et al. \cite{moise2016tracking} explored the use of Big Data analytics for tracking language mobility in geolocalised Twitter data. The authors proposed a two-step quantitative analysis. First, the authors studied the temporal evolution of languages on Twitter by investigating three case studies. Then, the subject of the study shifts from the temporal to the spatial perspective by observing the languages spread through monthly snapshots. Finally, starting from the insights obtained through temporal and spatial analyses, they study how linguistic mobility is reflected in the Twitter data. The insights obtained by the authors show that the adoption of Twitter is heterogeneous over time and reflects the linguistic distribution typical of the countries with a multilingual panorama (e.g., Switzerland).

Finally, Lamanna et al. \cite{lamanna2018immigrant} proposed an extensive study on the integration of immigrants using the Twitter language. First, the authors characterized immigrants through digital space-time communication patterns and then inferred immigrants' places of residence and native languages. Then, a modified entropy metric was introduced to quantify the so-called Power of Integration of cities - i.e., its ability to integrate different cultures - and to characterize the relationships between different cultures. While underlining some of the well-known concerns related to the use of Twitter data such as bias, the study seems to testify to the possibility of identifying spatial patterns and mobility profiles using this type of data.

While previous works concentrate on one dimension of OSNs in general and Twitter data in particular, our framework combines linguistic analysis, sentiment analysis and spatio-temporal information from geolocalized tweets. Moreover, our work includes two different languages, as opposed to previous works which generally concentrate on one language only, typically English. An important feature of our superdiversity index is the comparison with immigration rates and the high correlations observed, comparison which in previous works is not present or correlations are less relevant.

\section{Conclusions}
\label{sec:conclusions}
In this paper we proposed the novel Superdiversity Index ($SI$) that quantifies diversity in a population based on the changes in emotional content of words, compared to the standard language. We estimated this change in communities from the UK and Italy, using geolocalised Twitter data in English and Italian, at various geographical resolutions. In a majority of the geographical regions analysed we observed a remarkable correlation with foreign immigration rates, extracted from the dataset available from the Joint Research Center of the European Community, through the D4I data challenge. The proposed Superdiversity Index greatly outperforms other possible measures of diversity form the same Twitter data.

The method has been tested here for two countries and languages, following the nationality of the authors, but we plan to extend the analysis to other European countries. This work paves the way for a novel nowcasting model of immigration rates, that can be applied with higher time and space resolution compared to official statistics. In future work we plan to investigate the use of machine learning to achieve this task. We expect our superdiversity index to be a major feature in this model, with various other features used to correct for range differences, including language entropy, local dialects and population density.


\bibliographystyle{spmpsci}      


\begin{thebibliography}{10}
\providecommand{\url}[1]{{#1}}
\providecommand{\urlprefix}{URL }
\expandafter\ifx\csname urlstyle\endcsname\relax
  \providecommand{\doi}[1]{DOI~\discretionary{}{}{}#1}\else
  \providecommand{\doi}{DOI~\discretionary{}{}{}\begingroup
  \urlstyle{rm}\Url}\fi

\bibitem{aswad2018refugee}
Aswad, F.M.S., Menezes, R.: Refugee and immigration: Twitter as a proxy for
  reality.
\newblock In: The Thirty-First International Flairs Conference (2018)

\bibitem{barslund2016mobile}
Barslund, M., Busse, M.: How mobile is tech talent? a case study of it
  professionals based on data from linkedin.
\newblock A Case Study of It Professionals Based on Data from LinkedIn (June
  30, 2016). CEPS Special Report (140) (2016)

\bibitem{bradley1999affective}
Bradley, M.M., Lang, P.J.: Affective norms for english words (anew):
  Instruction manual and affective ratings.
\newblock Tech. rep., Citeseer (1999)

\bibitem{castellano2009nonlinear}
Castellano, C., Mu{\~n}oz, M.A., Pastor-Satorras, R.: Nonlinear q-voter model.
\newblock Physical Review E \textbf{80}(4), 041129 (2009)

\bibitem{d4i}
Centre, E.J.R.: Data challenge on integration of migrants in cities (2018).
\newblock \urlprefix\url{https://bluehub.jrc.ec.europa.eu/datachallenge/}

\bibitem{coletto2017perception}
Coletto, M., Esuli, A., Lucchese, C., Muntean, C.I., Nardini, F.M., Perego, R.,
  Renso, C.: Perception of social phenomena through the multidimensional
  analysis of online social networks.
\newblock Online Social Networks and Media \textbf{1}, 14--32 (2017)

\bibitem{Dubois2018}
Dubois, A., Zagheni, E., Garimella, K., Weber, I.: {Studying Migrant
  Assimilation Through Facebook Interests}  (2018)

\bibitem{esuli2007sentiwordnet}
Esuli, A., Sebastiani, F.: Sentiwordnet: a high-coverage lexical resource for
  opinion mining.
\newblock Evaluation pp. 1--26 (2007)

\bibitem{fiorio2017using}
Fiorio, L., Abel, G., Cai, J., Zagheni, E., Weber, I., Vinu{\'e}, G.: Using
  twitter data to estimate the relationship between short-term mobility and
  long-term migration.
\newblock In: Proceedings of the 2017 ACM on Web Science Conference, pp.
  103--110. ACM (2017)

\bibitem{freire2019characterization}
Freire-Vidal, Y., Graells-Garrido, E.: Characterization of local attitudes
  toward immigration using social media.
\newblock In: Companion Proceedings of The 2019 World Wide Web Conference, pp.
  783--790. ACM (2019)

\bibitem{gonccalves2014crowdsourcing}
Gon{\c{c}}alves, B., S{\'a}nchez, D.: Crowdsourcing dialect characterization
  through twitter.
\newblock PloS one \textbf{9}(11), e112074 (2014)

\bibitem{guerini2013sentiment}
Guerini, M., Gatti, L., Turchi, M.: Sentiment analysis: How to derive prior
  polarities from sentiwordnet.
\newblock arXiv preprint arXiv:1309.5843  (2013)

\bibitem{hawelka2014geo}
Hawelka, B., Sitko, I., Beinat, E., Sobolevsky, S., Kazakopoulos, P., Ratti,
  C.: Geo-located twitter as proxy for global mobility patterns.
\newblock Cartography and Geographic Information Science \textbf{41}(3),
  260--271 (2014)

\bibitem{HerdagdelenAmac2016}
{Herdagdelen, Ama{\c{c}}}, State, B., Adamic, L., Mason, W.: {The social ties
  of immigrant communities in the United States}.
\newblock In: WebSci (2016)

\bibitem{hu2004mining}
Hu, M., Liu, B.: Mining and summarizing customer reviews.
\newblock In: Proceedings of the tenth ACM SIGKDD international conference on
  Knowledge discovery and data mining, pp. 168--177 (2004)

\bibitem{hubl2017analyzing}
H{\"u}bl, F., Cvetojevic, S., Hochmair, H., Paulus, G.: Analyzing refugee
  migration patterns using geo-tagged tweets.
\newblock ISPRS International Journal of Geo-Information \textbf{6}(10), 302
  (2017)

\bibitem{kim2020digital}
Kim, J., S{\^\i}rbu, A., Giannotti, F., Gabrielli, L.: Digital footprints of
  international migration on twitter.
\newblock In: International Symposium on Intelligent Data Analysis, pp.
  274--286. Springer (2020)

\bibitem{lamanna2018immigrant}
Lamanna, F., Lenormand, M., Salas-Olmedo, M.H., Romanillos, G.,
  Gon{\c{c}}alves, B., Ramasco, J.J.: Immigrant community integration in world
  cities.
\newblock PloS one \textbf{13}(3), e0191612 (2018)

\bibitem{liu2005opinion}
Liu, B., Hu, M., Cheng, J.: Opinion observer: analyzing and comparing opinions
  on the web.
\newblock In: Proceedings of the 14th international conference on World Wide
  Web, pp. 342--351 (2005)

\bibitem{lyding2014paisa}
Lyding, V., Stemle, E., Borghetti, C., Brunello, M., Castagnoli, S.,
  Dell'Orletta, F., Dittmann, H., Lenci, A., Pirrelli, V.: The paisa'corpus of
  italian web texts.
\newblock In: 9th Web as Corpus Workshop (WaC-9)@ EACL 2014, pp. 36--43. EACL
  (European chapter of the Association for Computational Linguistics) (2014)

\bibitem{magdy2014exploiting}
Magdy, A., Ghanem, T.M., Musleh, M., Mokbel, M.F.: Exploiting geo-tagged tweets
  to understand localized language diversity.
\newblock In: Proceedings of Workshop on Managing and Mining Enriched
  Geo-Spatial Data, p.~2. ACM (2014)

\bibitem{Messias2016}
Messias, J., Benevenuto, F., Weber, I., Zagheni, E.: {From migration corridors
  to clusters: The value of Google+ data for migration studies}.
\newblock In: 2016 IEEE/ACM International Conference on Advances in Social
  Networks Analysis and Mining (ASONAM), pp. 421--428. IEEE (2016)

\bibitem{mocanu2013twitter}
Mocanu, D., Baronchelli, A., Perra, N., Gon{\c{c}}alves, B., Zhang, Q.,
  Vespignani, A.: The twitter of babel: Mapping world languages through
  microblogging platforms.
\newblock PloS one \textbf{8}(4), e61981 (2013)

\bibitem{moise2016tracking}
Moise, I., Gaere, E., Merz, R., Koch, S., Pournaras, E.: Tracking language
  mobility in the twitter landscape.
\newblock In: Data Mining Workshops (ICDMW), 2016 IEEE 16th International
  Conference on, pp. 663--670. IEEE (2016)

\bibitem{nielsen2011new}
Nielsen, F.{\AA}.: A new anew: Evaluation of a word list for sentiment analysis
  in microblogs.
\newblock arXiv preprint arXiv:1103.2903  (2011)

\bibitem{pollacci2017sentiment}
Pollacci, L., S{\^\i}rbu, A., Giannotti, F., Pedreschi, D., Lucchese, C.,
  Muntean, C.I.: Sentiment spreading: An epidemic model for lexicon-based
  sentiment analysis on twitter.
\newblock In: Conference of the Italian Association for Artificial
  Intelligence, pp. 114--127. Springer (2017)

\bibitem{riloff2003learning}
Riloff, E., Wiebe, J.: Learning extraction patterns for subjective expressions.
\newblock In: Proceedings of the 2003 conference on Empirical methods in
  natural language processing, pp. 105--112 (2003)

\bibitem{rodriguez2014migration}
Rodriguez, M., Helbing, D., Zagheni, E., et~al.: Migration of professionals to
  the us.
\newblock In: International Conference on Social Informatics, pp. 531--543.
  Springer (2014)

\bibitem{schmid1994probabilistic}
Schmid, H.: Probabilistic part-of-speech tagging using decision trees. in
  proceedings of international conference on new methods in language
  processing,(pp. 1-9), access date: 09.11. 2012 (1994)

\bibitem{schmid1995improvements}
Schmid, H.: Improvements in part-of-speech tagging with an application to
  german.
\newblock In: In proceedings of the acl sigdat-workshop. Citeseer (1995)

\bibitem{senaratne2014moving}
Senaratne, H., Br{\"o}ring, A., Schreck, T., Lehle, D.: Moving on twitter:
  using episodic hotspot and drift analysis to detect and characterise spatial
  trajectories.
\newblock In: Proceedings of the 7th ACM SIGSPATIAL International Workshop on
  Location-Based Social Networks, pp. 23--30 (2014)

\bibitem{sirbu2019}
S{\^\i}rbu, A., Andrienko, G., Andrienko, N., Boldrini, C., Conti, M.,
  Giannotti, F., Guidotti, R., Bertoli, S., Kim, J., Muntean, C.I., et~al.:
  Human migration: the big data perspective.
\newblock International Journal of Data Science and Analytics pp. 1--20 (2020)

\bibitem{sirbu2017opinion}
S{\^\i}rbu, A., Loreto, V., Servedio, V.D., Tria, F.: Opinion dynamics: models,
  extensions and external effects.
\newblock In: Participatory sensing, opinions and collective awareness, pp.
  363--401. Springer (2017)

\bibitem{socher2013recursive}
Socher, R., Perelygin, A., Wu, J., Chuang, J., Manning, C.D., Ng, A.Y., Potts,
  C.: Recursive deep models for semantic compositionality over a sentiment
  treebank.
\newblock In: Proceedings of the 2013 conference on empirical methods in
  natural language processing, pp. 1631--1642 (2013)

\bibitem{spyratos2018migration}
Spyratos, S., Vespe, M., Natale, F., Weber, I., Zagheni, E., Rango, M.:
  Migration data using social media: a european perspective  (2018)

\bibitem{state2013studying}
State, B., Weber, I., Zagheni, E.: Studying inter-national mobility through ip
  geolocation.
\newblock In: Proceedings of the sixth ACM international conference on Web
  search and data mining, pp. 265--274 (2013)

\bibitem{steiger2015advanced}
Steiger, E., De~Albuquerque, J.P., Zipf, A.: An advanced systematic literature
  review on spatiotemporal analyses of t witter data.
\newblock Transactions in GIS \textbf{19}(6), 809--834 (2015)

\bibitem{stewart2019rock}
Stewart, I., Flores, R.D., Riffe, T., Weber, I., Zagheni, E.: Rock, rap, or
  reggaeton?: Assessing mexican immigrants' cultural assimilation using
  facebook data.
\newblock In: The World Wide Web Conference, pp. 3258--3264. ACM (2019)

\bibitem{stone1966general}
Stone, P.J., Dunphy, D.C., Smith, M.S.: The general inquirer: A computer
  approach to content analysis.  (1966)

\bibitem{taboada2011lexicon}
Taboada, M., Brooke, J., Tofiloski, M., Voll, K., Stede, M.: Lexicon-based
  methods for sentiment analysis.
\newblock Computational linguistics \textbf{37}(2), 267--307 (2011)

\bibitem{templin1957certain}
Templin, M.C.: Certain language skills in children; their development and
  interrelationships.  (1957)

\bibitem{vertovec2007super}
Vertovec, S.: Super-diversity and its implications.
\newblock Ethnic and racial studies \textbf{30}(6), 1024--1054 (2007)

\bibitem{wilson2005recognizing}
Wilson, T., Wiebe, J., Hoffmann, P.: Recognizing contextual polarity in
  phrase-level sentiment analysis.
\newblock In: Proceedings of human language technology conference and
  conference on empirical methods in natural language processing, pp. 347--354
  (2005)

\bibitem{zagheni2014inferring}
Zagheni, E., Garimella, V.R.K., Weber, I., et~al.: Inferring international and
  internal migration patterns from twitter data.
\newblock In: Proceedings of the 23rd International Conference on World Wide
  Web, pp. 439--444. ACM (2014)

\bibitem{zagheni2012you}
Zagheni, E., Weber, I.: You are where you e-mail: using e-mail data to estimate
  international migration rates.
\newblock In: Proceedings of the 4th Annual ACM Web Science Conference, pp.
  348--351. ACM (2012)

\bibitem{zagheni2017leveraging}
Zagheni, E., Weber, I., Gummadi, K., et~al.: Leveraging facebook’s
  advertising platform to monitor stocks of migrants.
\newblock Population and Development Review \textbf{43}(4), 721--734 (2017)

\end{thebibliography}

\end{document}